\documentclass[12pt,draftclsnofoot,onecolumn]{IEEEtran}

\usepackage{amsmath, mathrsfs, mathtools}
\usepackage{amsfonts}
\usepackage{amssymb}
\usepackage{amsthm}
\usepackage{dsfont}
\usepackage{graphicx}
\usepackage{hyperref}
\usepackage[caption=false]{subfig}
\usepackage{xcolor}
\usepackage[backend=biber,giveninits=true,sorting=none]{biblatex}
\renewcommand*{\bibfont}{\footnotesize}

\long\def\comment#1{}

\newcommand{\ben}{\begin{enumerate}}
\newcommand{\een}{\end{enumerate}}

\newcommand{\beq}{\begin{equation}}
\newcommand{\eeq}{\end{equation}}

\newcommand{\bi}{\begin{itemize}}
\newcommand{\ei}{\end{itemize}}

\DeclareMathOperator*{\argmax}{arg\,max}
\newcommand{\norm}[1]{\lVert#1\rVert}

\newcommand{\CC}{\mathbb{C}}
\newcommand{\PP}{\mathbb{P}}
\newcommand{\RR}{\mathbb{R}}

\newcommand{\FF}{\mathbb{F}}

\newcommand{\EE}{\mathbb{E}}

\newcommand{\av}{{\bf a}}
\newcommand{\bv}{{\bf b}}
\newcommand{\cv}{{\bf c}}

\newcommand{\ev}{{\bf e}}
\newcommand{\fv}{{\bf f}}
\newcommand{\gv}{{\bf g}}
\newcommand{\hv}{{\bf h}}

\newcommand{\pv}{{\bf p}}
\newcommand{\qv}{{\bf q}}
\newcommand{\rv}{{\bf r}}
\newcommand{\sv}{{\bf s}}

\newcommand{\uv}{{\bf u}}
\newcommand{\wv}{{\bf w}}
\newcommand{\vv}{{\bf v}}
\newcommand{\xv}{{\bf x}}
\newcommand{\yv}{{\bf y}}
\newcommand{\zv}{{\bf z}}
\newcommand{\zerov}{{\bf 0}}
\newcommand{\onev}{{\bf 1}}

\newcommand{\Am}{{\bf A}}
\newcommand{\Bm}{{\bf B}}
\newcommand{\Cm}{{\bf C}}

\newcommand{\Fm}{{\bf F}}

\newcommand{\Hm}{{\bf H}}
\newcommand{\Id}{{\bf I}}

\newcommand{\Pm}{{\bf P}}

\newcommand{\Rm}{{\bf R}}

\newcommand{\Um}{{\bf U}}
\newcommand{\Wm}{{\bf W}}

\newcommand{\Xm}{{\bf X}}
\newcommand{\Ym}{{\bf Y}}
\newcommand{\Zm}{{\bf Z}}

\newcommand{\lambdav}{\hbox{\boldmath$\lambda$}}
\newcommand{\epsilonv}{\hbox{\boldmath$\epsilon$}}

\newcommand{\muv}{\hbox{\boldmath$\mu$}}
\newcommand{\zetav}{\hbox{\boldmath$\zeta$}}
\newcommand{\phiv}{\hbox{\boldmath$\phi$}}
\newcommand{\psiv}{\hbox{\boldmath$\psi$}}

\newcommand{\Psim}{\boldsymbol{\Psi}}

\newcommand{\trace}{{\hbox{tr}}}

\newcommand{\SNR}{{\sf SNR}}

\newcommand{\herm}{{\sf H}}

\theoremstyle{remark}
\newtheorem{theorem}{\bf Theorem}

\newtheorem{remark}{\bf Remark}

\newtheorem{assumption}{\bf Assumption}

\title{Sparse Regression LDPC Codes for the Block-Fading Non-Coherent SIMO Channel}
\author{Alexander Fengler, Burak \c{C}akmak, Giuseppe Caire
\thanks{
A. Fengler was funded by the Deutsche Forschungsgemeinschaft (DFG, German Research Foundation) – Grant 471512611. 
The work of B. \c{C}akmak was supported by the Gottfried Wilhelm Leibniz-Preis 2021 of DFG. The work of G. Caire was supported by BMBF Germany in the program of ``Souver\"an. Digital. Vernetzt.'' Joint Project 6G-RIC (Project IDs 16KISK030).
A. Fengler is with the Department of Electrical Engineering and Information Technology, Karlsruhe Institute of Technology, 76131 Karlsruhe, Germany.
B. \c{C}akmak and G. Caire are with the Faculty of Electrical Engineering and Computer Science, Technical University of  Berlin, 10587 Berlin, Germany.
Authors emails: fengler@kit.edu,\{burak.cakmak,caire\}@tu-berlin.de.}}
\begin{document}

\maketitle
\begin{abstract}
Sparse regression codes (SPARCs) are a class of codes that encode information through the superposition of columns of a randomised coding matrix. The combination with an outer non-binary low density parity check (NB-LDPC) code was recently shown to improve the finite-length performance of these codes over the unfaded AWGN channel. In this paper, we propose a low-complexity approximate message passing (AMP) decoder that is capable of decoding NB-LDPC encoded SPARCs on a Rayleigh fading channel with multiple receive antennas. Notably, the decoder does not require channel state information (CSI), i.e., it is fully non-coherent, but achieves the same error probability as a decoder with full CSI, even for moderate block lengths. This is achieved by iteratively re-estimating the channel throughout the decoding iterations. In addition, we provide a rigorous asymptotic analysis of both the block error probability and the channel estimation error. Numerical results confirm the precision of the analysis and show that the presented coding scheme performs within 1.5 dB of the outage capacity and is competitive with coded modulation schemes employing standardised LDPC codes for 5G cellular networks and pilot-based channel estimation.   
\end{abstract}


\section{Introduction}
\label{sec:intro}
We consider the problem of coding for a single-input frequency-non-selective slowly fading channel with multiple receive antennas. 
\begin{equation}
    \Ym = \zv\hv^\herm + \Wm,\quad \zv\in\CC^n,\hv\in\CC^M
    \label{eq:channel}
\end{equation} 
where $\zv$ represents the encoded and modulated message and $\hv$ contains the complex conjugate of the channel coefficients for each of the $M$ receive antenna.
The model \eqref{eq:channel} is also known as the quasi-static fading single-input multiple-output (SIMO) channel \cite{Big98,Yan14}. 
This model is relevant for narrow-band applications with latency constraints and low mobility, e.g., real-time control or sensor networks. In this setting, the transmission time is shorter than the coherence block length
\footnote{The coherence block length is approximately  the product of the coherence time and the coherence bandwidth of the channel, and indicates the number of time-frequency signal dimensions over which the channel coefficients can be considered constant.}
of the channel and averaging over multiple fading blocks is not possible. 
Compared to an additive white Gaussian noise (AWGN) channel, the effective received SNR at each antenna is a random variable that depends on the channel realisation. Hence, the block error probability does not vanish with increasing block length, and therefore the Shannon capacity is zero. A more meaningful performance measure is the information outage probability $\PP(R<\log(1+\|\hv\|^2\SNR))$, where $\SNR$ denotes the transmit SNR, which corresponds to the best achievable block error rate (BLER) for a fixed coding rate $R$ when channel state information (CSI) is available at the receiver \cite{Big98}. It is shown in \cite{Yan14} that the channel dispersion for quasi-static fading channels is zero and the finite-length BLER is sharply dominated by the information outage probability, which is therefore the ``right’’ proxy for BLER also under a more refined analysis.
In principle, the channel can be estimated by sending pilot symbols. The receiver then equalises the signal, effectively converting it to a Gaussian channel with $\SNR_\text{eff} = \|\hv\|^2\SNR$. However, this approach has some major downsides. 
For once, the CSI obtained from pilot symbols is noisy, and the conventional quasi-coherent receivers that extract the CSI from pilots and then treat it as if it is perfect are inherently suboptimal. Furthermore, in the uplink, users transmit with low power, resulting in noisy channel estimates, unless one devotes a significant fraction of power to the pilots. Since the user's transmit power is peak limited, it is impossible to transmit a single pilot symbol with a huge power. This makes it necessary to transmit multiple pilot symbols, which incurs an overhead in the time-frequency resources.   
Even for moderated block-lengths of 3-4 thousand the overhead created by pilot symbols is non-negligible.
In a multi-user scenario, the situation is even worse, and pilot contamination is a huge source of degradation when pilots from different users are not orthogonal \cite{Mar10}. Hence, in massive MIMO, the pilot dimension must grow with the number of users served in the uplink on each time-frequency slot. Since the number of users $K$ grows linearly with the number of receive antennas $M$ (although, with a smaller factor, typically $M/K = 10$), this means that for larger and larger systems, and fixed coherence block length, at some point the pilot dimension in the multiuser case incurs a major overhead. Overall, a pilot-less transmission scheme is generally highly desirable, and even if we only treat the single-user case in this paper, it is to be seen as a step towards developing multi-user coding schemes that can be efficiently decoded without CSI.

Sparse Regression Codes (SPARCs) have been proposed by Joseph and Barron as a new class of error correcting codes for the AWGN channel \cite{Jos12c}. For encoding, the message is first represented by a sequence of position modulated symbols (also called sections). The position indices in each section are used to select columns from a sectionised base matrix, which are then superposed to create the final encoder output. In particular, the base matrix is over the reals \footnote{In this paper, we use a straightforward generalisation to complex valued codebooks.},  and the empirical distribution of the codewords is Gaussian, matching the capacity achieving input distribution. This alleviates the need for further modulation or shaping.
SPARCs have been shown to achieve the capacity of the AWGN channel with a polynomial encoding complexity and maximum likelihood decoding \cite{Jos12c}. Later, it was shown that capacity can be achieved with low-complexity approximate message passing (AMP) decoding if power allocation or spatial coupling is used \cite{Rus17b, Bar17}, resulting in a capacity-achieving coding scheme for the AWGN channel with polynomial encoding and decoding complexity.
It was already noted in \cite{Jos12c} that a vanishing section error rate
can be achieved at rates up to capacity, but not a vanishing BLER. That requires the use of an outer code with a rate approaching one. Yet, it was noted in \cite{Gre18} that a lower rate outer code can improve the finite-length behaviour of SPARCs. The decoding of the inner and outer codes in \cite{Gre18} was separated. In a parallel line of work \cite{Liu21} Ping et. al. showed that compressed sensing based coding schemes can achieve capacity through the use of a properly designed outer code if inner and outer code are decoded iteratively. 
In \cite{Ebe25} a concatenated coding scheme of SPARC and non-binary low density parity check (NB-LDPC) code, termed sparse-regression LDPC (SR-LDPC),  was presented together with an iterative decoder that combines AMP with classical LDPC belief propagation. 

AMP was introduced in \cite{kabashima2003cdma,Don09} as an iterative reconstruction algorithm for linear inverse problems. In \cite{Bay11} it was shown that the mean square reconstruction error can be rigorously tracked in the asymptotic limit of large dimensions by the state evolution (SE) formalism, a set of recursive equations. Several generalisations of the base algorithm were developed, including generalised AMP (GAMP) which handles separable but non-Gaussian output channels \cite{Ran11}, MMV-AMP which handles linear inverse 
\emph{matrix} recovery problems \cite{Kim11}, and vector/orthogonal AMP (VAMP/OAMP) which extend the algorithm to a wider class of sampling matrices \cite{Ma17,Ran19}. Recent results also established concentration of the reconstruction error around the SE for VAMP/OAMP and its generalised versions \cite{Cad24,Cak25b}. Another class of AMP algorithms is able to solve bilinear recovery problems of the form $\Ym = \Um\Hm^\herm + \Wm$ where both $\Um$ and $\Hm$ are unknown and have columns from known prior distributions $p_\hv,p_\xv$ \cite{Par14a,Fle18,Mon21}.  

Our main contributions are as follows.
We present an extension of the SR-LDPC coding scheme for the Rayleigh slow-fading channel with multiple receive antennas. 
Specifically, we present an AMP algorithm that is capable of jointly decoding the SR-LDPC code and recovering the channel vector. Notably, we find that no loss is incurred due to the lack of CSI at the receiver. That is, the proposed decoder achieves, without transmission of pilot symbols, the same BLER as an oracle version of the decoder that has full CSI. It can do so with a marginal computational complexity overhead. We  give a fully rigorous asymptotic analysis that accurately predicts the BLER and channel estimation error, allowing for a quick estimation of BLER without the need for extensive Monte-Carlo simulations. 
  
Note that even trivial channel estimation techniques, such as averaging the received signal over the time slot, can provide an asymptotically exact channel estimate in the limit of large dimensions. However, for the moderate length regime ($<$ 10k) our proposed algorithm leads to a significantly lower channel estimation error due to the joint estimation of message and channel.

\subsection{Related Work}
Multi-user versions of SPARCs have been developed in \cite{Fen21e,Ama22}.
In \cite{Ago23}, a bilinear joint decoding and channel estimation algorithm for multi-user SPARCs was proposed. The approach in \cite{Ago23} uses the grouping \eqref{eq:outer_mmv}, explained in more detail in Section \ref{sec:decoding}. That is, the bilinear AMP is used to recover the matrix $\xv\hv^\herm$ from the noisy outer product \eqref{eq:channel}, leading to a complexity $\mathcal{O}(MN\log N)$, where $N = LB$ with $L$ being the number of sections of the SPARC and $B$ their size. In contrast, our approach applies to the grouping \eqref{eq:outer1}, leading to a complexity  of $\mathcal{O}(N \log N + nM)$ operations per iteration. This results in a significant reduction in complexity of a factor $M$ and in a more stable algorithm.   

The problem itself is a bilinear recovery problem with a simple (Gaussian) prior on $\hv$ and a complicated (SPARC) prior on $\zv$. So in principle it could be solved by bilinear AMP. However, due to the complicated prior on $\zv$ the estimation of $\zv$ is highly non-trivial. That’s why we resort to a different approach: We treat the problem as a linear inverse recovery problem with an output channel $\zv \mapsto \zv\hv^\herm + \Wm$ and a postulated output channel $\zv \mapsto \zv\hv_t^\herm + \Wm$. The parameter $\hv_t$ is iteratively updated throughout the iterations $t$. This results in the proposed algorithm and allows for analysis by considering the SE of a mismatched posterior mean estimator.

A novel maximum likelihood matching pursuit decoder (MLMP) for SPARCs on Rayleigh fading SIMO channels was recently proposed in \cite{Kan25}. The decoder was an extension of a greedy matching decoder which was shown to outperform AMP for short blocklength ($n=128$ complex channel uses) on the AWGN channel \cite{Sin24}. In contrast, our work targets intermediate blocklengths of several thousand. At these lengths the MLMP decoder becomes impractical since it is necessary to store the whole dictionary to compute terms like Eqs. (18) or (19) in \cite{Kan25}. An implementation of MLMP that can use the properties of structured matrices to avoid storage of the entire dictionary was not known at the time of writing. Furthermore, it was shown in \cite{Ebe25} that at high rates an outer code with iterative decoding is necessary to optimise $E_b/N_0$ for SPARCs which is not possible with the existing versions of MLMP.

\section{Encoding}
\label{sec:encoding}
A SPARC is defined by the number of sections $L$, the section size $B$, the blocklength $n$, and base dictionary $\Am$ of size $n\times BL$. We also define $N = BL$. 

For an SR-LDPC code, the SPARC is concatenated with an outer NB-LDPC code whose alphabet size and length corresponds to $B$ and $L$, respectively.
The outer code is defined by a parity check matrix $\Hm \in \FF_B^{(L-L_\text{inf})\times L}$ with elements in the Galois field $\FF_B = \text{GF}(B)$.
The set of codewords are all $\xv \in \FF_B^{L}$ that satisfy $\Hm \xv = \zerov$. We assume that $\Hm$ has full rank, resulting in $B^{L_\text{inf}}$ distinct codewords. The elements of $\FF_B$ are identified with unit basis vectors $\lambdav_i \in \{0,1\}^B$, with $\lambda_{ii} = 1$ and $ \lambda_{ij} = 0\ \forall j\neq i$, by the mapping $q \mapsto \lambdav_q$. For further details, see \cite{Ebe25}. 
For example, consider the outer non-binary LDPC code as in \cite{Ebe25} with $L=766$ sections of size $B = 2^8$, and $L_\text{inf} = 736$ information sections. For SR-LDPC encoding, a $b = 2^{L_\text{inf}\log_2 B}$ bit message is first encoded by the outer NB-LDPC code into $L$ symbols $(i_1,...,i_L)$ with $i_j \in [B]$. The final encoder output is $\zv = \sqrt{nP}\sum_{j=1}^L\av_{i_j}$. The base dictionary is normalised so that $\EE[\|\zv\|^2] = nLP$, where we use $\|\cdot\|:=\|\cdot\|_2$ to denote the 2-norm. 
The rate of the outer NB-LDPC code is $R_\text{out} = b/(L\log_2 B) = L_\text{inf}/L$ and the rate of the inner SPARC is $R_\text{in} = (L\log_2 B)/n$.
Accordingly, the total code rate is $R = R_\text{in}R_\text{out} = b/n$.

The noise matrix $\Wm$ has i.i.d. Gaussian entries with zero mean and variance \(\sigma^2\), denoted as
\(W_{ij} \sim_{\text{i.i.d.}} \mathcal{CN}(0, \sigma^2)\).
The transmit SNR is then 
$  \SNR = \frac{LP}{\sigma^2}$
and the transmit energy per bit is 
$
\frac{E_b}{N_0} = \frac{\SNR}{R}$.  Furthermore, while the proposed framework is valid for any distribution of the channel coefficients, 
the numerical results are based on a Rayleigh fading model, 
i.e., \(h_j \sim_{\text{i.i.d.}}\mathcal{CN}(0,1)\).

In this paper, we use randomly signed subsampled discrete Fourier transform (DFT) matrices \cite{And14,wang2024universality,Dudeja24} as SPARC dictionaries. Formally, we have

\begin{equation}
\Am= \sqrt{\frac N n}\Pm \Fm_{N}{\rm diag}(\sv)     
\label{eq:codebook}
\end{equation}
where $\Pm\in\{0,1\}^{n\times N}$ is a diagonal projection matrix that randomly selects $n$ rows from the (unitary) DFT matrix  $\Fm_{N}$, and $\sv\in\{\pm 1\}^{N}$ is uniform i.i.d..
Let $\xv \in \RR^{N}$, with $\xv = \sqrt{nP}[\lambdav_{i_1}^\top,...,\lambdav_{i_L}^\top]^\top$, be the sparse vector that represents the outer encoded message with $x_{i_j + jB} = \sqrt{nP}$ and zero otherwise. With this, the final encoded message is $\zv = \Am \xv$. The DFT matrix allows to compute $\Am\xv$ and $\Am^\herm \yv$ via fast Fourier transforms in $\mathcal{O}(N\log N)$ operations.
\begin{remark}
 The random $\pm 1$ sequence eliminates possible unwanted correlations between the columns of the codebook \cite{And14}. This phenomenon is most notably observed when trying to compute $\mathds{1}^\top\Am^\herm\Ym$, which we will use later for channel estimation. Without random signage,  due to basic properties of the Fourier transform,  $\mathds{1}^\top\Am^\herm$ is a vector whose only nonzero is in the first entry, leading to a useless estimate. 
\end{remark}

\section{Decoding}
\label{sec:decoding}

\subsection{A high level overview}
The channel model \eqref{eq:channel} with SPARC encoding can be parenthesised in two ways:
\begin{equation}
    \Ym = (\Am\xv)\hv^\herm + \Wm
    \label{eq:outer1}
\end{equation}
or
\begin{equation}
    \Ym = \Am(\xv\hv^\herm) + \Wm.
    \label{eq:outer_mmv}
\end{equation}

We briefly discuss the advantages and disadvantages of the two equivalent formulations.
The second formulation \eqref{eq:outer_mmv} leads to an MMV recovery problem, where the unknown is the matrix $\xv\hv^\herm$. Recent works that design an AMP algorithm for similarly structured MMV recovery problems include \cite{Fen22c,Liu18}. The downside of the MMV approach is that the resulting AMP algorithms have a high per-iteration complexity of order $\mathcal{O}(NM\log N)$ simply because terms like $\Am(\xv_t\hv_t^\herm)$ and $\Am^\herm\Ym_t$ need to be computed every iteration, where $\xv_t,\hv_t,\Ym_t$ denote iterative estimates of the associated quantities. 

In this paper, we describe two variations of an AMP algorithm based on formulation \eqref{eq:outer1}. Since $\hv$ is not known, we treat it as a fixed but unknown parameter and estimate it as part of the AMP iterations. The resulting algorithm first estimates $\zv$ and then runs a standard single-measurement vector AMP iteration to estimate $\xv$ instead of trying to detect the nonzero rows of the matrix $\xv\hv^\herm$. This approach has a lower complexity and does not suffer from divergence issues sometimes observed in MMV-AMP \cite{Fen21h}.

The first variation we give is derived from the standard AMP framework, while the second variation is based on the orthogonal AMP (OAMP) \cite{Ma17} or vector AMP (VAMP) \cite{Ran19} framework. OAMP and VAMP describe the same algorithm, and we refer to it as OAMP throughout this paper. It is known that standard AMP can show unstable behaviour in certain situations, see \cite{Ma17}. Such behaviour can be circumvented by OAMP. Furthermore, OAMP takes into account the partially orthogonal nature of the dictionary matrix $\Am$ while AMP does not, resulting in a potentially better algorithm.

\subsection{AMP Algorithm}

Start with $t=0$, , $\pv_{0} = \zerov$, $\xv_{0} = \zerov$, $\hv_{0} = \hv_\text{init}$ and iterate:
\begin{subequations}
\begin{align}
\uv_{t}&=\frac{\Ym\hv_{t}}{\Vert\hv_{t} \Vert^2 }-\pv_{t}\\
\rv_{t}&=\xv_{t}+\Am^\herm\uv_{t}\\
\tau^2_{t}&=\frac{\|\uv_t\|^2}{n}\label{eq:amp_tau}\\
\xv_{t+1}&=\eta(\rv_{t},\tau_t) \label{eq:amp_s}\\
q_{t+1} &= \langle\eta'(\rv_{t},\tau_t) \rangle \\
\pv_{t+1}&=\Am\xv_{t+1}-\delta q_{t+1} \uv_{t}\label{eq:amp_p}\\
\hv_{t+1}&=\frac{ \Ym^\herm \pv_{t+1}}{\sigma^2+\Vert \pv_{t+1}\Vert^2}\label{eq:h_hat}
\;.
\end{align}   
\label{eq:amp}
\end{subequations}

Where $\delta = N/n$ denotes the oversampling ratio, and $\hv_\text{init}$ is an arbitrary initial channel estimate that will be discussed later in Section~\ref{sec:init}.
The denoising functions $\eta(\rv,\tau_t)$ are applied section-wise and will be discussed in the following Section~\ref{sec:denoising}. $\eta'(\rv,\tau_t)$ denotes the vector $(\partial \eta_{i}(\rv)/ \partial r_i)_{i=1,...,N}$ and $\langle f(\xv) \rangle := (\sum_i f_i(\xv))/N$ the empirical average for a function $f: \CC^N \to \CC^N$. We refer to $\langle \eta'(\rv,\tau) \rangle$ as the Onsager term. 
The decoder is run for $T$ iterations or until the following early stoppage criterion is met:
$\xv^t$ is first quantised to a valid SPARC signal $\hat \xv$, that is, in each section $\ell$ the value of $\hat \xv$ with index $\argmax_{i\in[B]} x^t_{\ell,i}$ is set to $1$ and all other entries in that section to $0$. 
If the resulting vector, converted to a vector over $\FF_B$, satisfies $\Hm\hat \xv = \zerov$, the decoder stops and outputs $\hat \xv$.

The form of the channel estimate in line \eqref{eq:h_hat} will be justified by our analysis later. However, an intuitive explanation is as follows. The MMSE estimate of $\hv$ would be given as
\begin{equation}
    \hat{\hv}^\text{MMSE} = \EE[\hv|\Ym] = \EE[\EE[\hv|\Ym,\zv] |\Ym]
\end{equation}
Since this integral is hard to compute, we can approximate it by a point estimate as $\EE[\hv|\Ym,\zv^*]$ with $\zv^* = \EE[\zv|\Ym]$. That in itself is not very useful, since $\EE[\zv|\Ym]$ is equally hard to compute. However, the AMP iterations provide us with an iterative estimate of $\EE[\zv|\Ym]$ in the form of $\pv_t$.  Calculating $\EE[\hv|\Ym,\zv = \pv_t]$ gives exactly \eqref{eq:h_hat}.

\subsection{OAMP Variant}
The OAMP version differs from \eqref{eq:amp} in that the lines \eqref{eq:amp_tau},\eqref{eq:amp_p}, and \eqref{eq:amp_s} are replaced by
\begin{subequations}
\begin{align}
\tau^2_{t}&=\frac{1}{\delta}\left[
\frac{\sigma^2}{\|\hv_{t}\|^2}+(\delta-1)\frac{\Vert\uv_t \Vert^2}{n}\right]\label{eq:oamp_tau}\\
\xv_{t+1}&=\frac{\eta(\rv_t,\tau_t) - q_{t+1}\rv_t}{1-q_{t+1}}\\
\pv_{t+1}&=\Am \xv_{t+1}\;.
\end{align}  
\label{eq:oamp}
\end{subequations} 

\subsection{Denoising Functions}
\label{sec:denoising}
We discuss here the design of the denoising function $\eta$ when $\xv \in \{0,1\}^{BL}$ consists of sections $\xv_\ell, \ell = 1,...,L$ of size $B$ chosen as unit vector representation of a $B$-ary codeword from some outer code. Since for most codes of interest the coded symbols are uniformly distributed, we can model their marginal distribution as $P(\xv_\ell = \sqrt{nP}\lambdav_j) = 1/B$ for $j=1,...,B$, where $\lambdav_j$ is a vector of zeros with a single one in position $j$. 

In Section~\ref{sec:proof} we verify that, in the large-system limit ($n, N \to \infty$), the input $\rv_t$ to the denoiser has the \emph{effective} (formally stated in Theorem~\ref{Th1} in Section~\ref{sec:proof}) distribution as
\begin{align}
    \rv_t \sim \mathcal{CN}(\alpha_t \xv,\tau_t^2\Id_N) \label{eq:r_gauss}
\end{align}
where we define for $t\geq 0$
\begin{equation}
    \alpha_{t} := \frac{\hv^\herm\hv_t}{\|\hv_t\|^2}
\end{equation}
The factor $\alpha_t$ in \eqref{eq:r_gauss} is caused by a mismatch between the true channel $\hv$ and its estimate $\hv_t$. In fact, our SE analysis in Section~\ref{sec:analysis} reveals that $1/\alpha_t$ is asymptotically given by the normalised overlap $\xv^\top \xv_{t}/\|\xv_{t}\|^2$ between the transmitted message and its prediction. However, $\alpha_t$ is generally unknown at the receiver, so we design the denoiser based on the matched model $\alpha_t = 1$. We note that it is possible to design a denoiser at each iteration that incorporates an algorithmic prediction of $\alpha_t$ (see Remark 3), although we do not address such a scheme in this paper.
 
We consider the following denoising function \cite{Ebe25}, which is a concatenation of the two functions $\eta_1$, applied section-wise (the section index $\ell$, and the iteration index $t$, are dropped for readability when not needed), and $\eta_2$ applied on all sections jointly:
\begin{equation}
    \eta(\rv,\tau) = \sqrt{nP}\eta_2\left(\eta_1\left(\rv,\frac{\tau}{\sqrt{2}}\right)\right)
    \label{eq:eta}
\end{equation}
where
\begin{equation}
    \eta_1(\rv, \tau) = \text{softmax}\left(\exp\left(\frac{\sqrt{nP}}{\tau^2}\rv\right)\right)
    \label{eq:eta1}
\end{equation}
with the softmax function 
\begin{equation}
    \text{softmax}(x_1,...,x_B)_i = \frac{e^{x_i}}{\sum_j e^{x_j}}.
    \label{eq:softmax}
\end{equation}  

The output of $\eta_1$ is precisely a vector of posterior probabilities $\PP(\xv_\ell = \sqrt{nP}\lambdav_j| \rv_\ell)$ in the Gaussian channel $\rv_\ell = \xv_\ell + \tau\wv_\ell$ under the uniform prior on $\xv$. The factor $1/\sqrt{2}$ in \eqref{eq:eta} is because the estimated signal is real, while $\rv$ is complex valued, and $\tau^2$ refers to the complex noise power. The vector of posterior probabilities is subsequently used as a priori probabilities for a soft-in soft-out (SISO) belief propagation (BP) decoder, which we denote by $\eta_2(\sv)$. 
The form of the BP denoiser can be found in, e.g., \cite{Ebe25}. We recall the definition here for completeness. 

Denote the section-wise probability vectors output by $\eta_1$ as $\muv^0_i \in \RR^B,\ i \in [L]$. We use $\mu_0(q),\ q\in [B]$ to represent its entries. Let $a \in [L - L_\text{inf}]$ denote the index of a row of the parity check matrix $\Hm$ of the outer code, and $i \in [L]$ denote the index of a column. We refer to those as check and variable nodes.
For a given number of SISO iterations the BP decoder iteratively computes message updates passed between check and variable nodes as:
\begin{align}
    \mu_{i \to a}(q) &\sim \mu_i^0(q)\prod_{b\in \partial i\setminus a} \mu_{b \to i}(q)\label{eq:bp_vntocn}\\
    \mu_{a\to i}(q) &\sim \sum_{\substack{\qv'\in \FF_B^{|\partial a\setminus i|}\\\Hm_{a,\partial a\setminus i}\qv' = - H_{a,i}q}} \prod_{j \in \partial a\setminus i} \mu_{j\to a}(q'_j)\label{eq:bp_cntovn}
\end{align}
Here $\partial a := \{i: H_{ai}\neq 0\}$ refers to the neighbourhood of check node $a$, with the analogous definition for variable nodes. For a set $S$, $\Hm_{a,S}$ refers to the subvector of the $a$-th row of $\Hm$ indexed by $S$. Both message updates are normalised to one after each step such that $\sum_{q\in [B]} \mu_{i\to a}(q) = \sum_{q\in [B]} \mu_{a\to i}(q) = 1$. The check-to-variable node update is an $|\partial a\setminus i|$-fold circular convolution of permuted versions of the vectors $\muv_{j\to a}$, which can be efficiently computed using a fast Hadamard transform in $\mathcal{O}(r_\text{max}B\log B)$ steps, where $r_\text{max}$ is the maximum check node degree. For more details, see \cite{Ebe25}. The final output of the denoiser for section $i$ after a fixed number of SISO iterations is 
\begin{equation}
    \eta(\rv,\tau)_i = \sqrt{nP}\frac{\muv_i^0\prod_{b\in \partial i} \muv_{b \to i}}{\|\muv_i^0\prod_{b\in \partial i} \muv_{b \to i}\|_1}\;,\label{eq:bp_out}
\end{equation}
where the product is entry-wise.
\subsubsection{Onsager Term}
As shown in \cite{Ebe25} the BP decoder $\eta_2$ is Lipschitz and does not contribute to the Onsager term $q_{t}$, if it is run for fewer iterations than the girth (length of the smallest cycle) of the defining Tanner graph. We will run the BP decoder only for one iteration, based on considerations in the following analysis section \ref{sec:siso}.
The derivative of $\eta_1$ is given as:
\begin{equation}
    \eta_1'(\rv,\tau) = \eta_1\left(\rv,\frac{\tau}{\sqrt{2}}\right)\odot\left[\onev-\eta_1\left(\rv,\frac{\tau}{\sqrt{2}}\right)\right]\frac{\sqrt{nP}}{\tau^2},
\end{equation}
where $\odot$ denotes the entry-wise (Hadamard) product of two vectors, $\onev$ is the all one vector, and $\eta'(\rv,\tau) = \sqrt{nP}\eta_1'(\rv,\tau)$.
\subsection{Initialization}
\label{sec:init}
Since the iterations are most conveniently initialised with $\xv_0 = \zerov$ a non-zero initial estimate of $\hv$ is required in the first iteration. Otherwise, the algorithm cannot get away from the all-zero fixed point. Random initialisation is possible, but generally suboptimal. We find that the following simple initialisation, which takes into account that $\xv$ is non-negative, leads to optimal reconstruction performance (in the sense that it allows us to reach the same performance as the decoder with full CSI).
\begin{equation}
    {\hv}_\text{init} = \frac{\Ym^\herm\Am\mathds{1}}{\|\xv\|_1}
    \label{eq:h_init}
\end{equation}
which is equivalent to averaging the rows of the matrix $\Am^\herm\Ym$.

\begin{remark}\label{initdecom}
Using basic properties of random matrices we show in Appendix \ref{prem1} that for a Gaussian random matrix ensemble with $A_{ij}\sim_{\text{i.i.d.}} \mathcal {CN}(0,1/n)$ we can express   
\begin{equation}\label{decom}
{\hv}_\text{init}= \left(\gv_0 \hv^\herm + \sqrt{\frac {1}{nPL}}\Wm\right)^\herm\left(\gv_0 +\sqrt{B-1}\gv_1\right)
\end{equation}
where $\gv_0\sim \mathcal {CN}(\mathbf {0},\Id_n/n)$, $\gv_1\sim \mathcal {CN}(\mathbf{0},\Id_n/n)$ and $\Wm$ are all mutually independent. For a right unitarily invariant matrix ensemble, which closely resembles \eqref{eq:codebook}, the same decomposition \eqref{decom} holds, but with a slightly different distribution for $\gv_0$ and $\gv_1$. Specifically, 
$\gv_i$ are distributed as $\sqrt{\delta} \Pm \uv_i/\| \uv_i \|$ where $\uv_i \sim \mathcal{CN}(\mathbf{0}, \Id_N)$ and $\Pm \in \{0,1\}^{n \times N}$ is a selection matrix that extracts $n$ entries from $\uv_i$. The derivation of \eqref{decom} in this case is technically more involved, and we omit it. Similar decomposition results can be found in \cite{lu2021householder,ccakmak2025multi}. 
\end{remark}

Informally, we note that for both matrix ensembles, 
$\mathbf{g}_i^\herm \mathbf{g}_j = \delta_{ij} + \mathcal{O}(n^{-1/2})$ for large $n$. 
Hence, the result implies that
\begin{equation}
    \left\| \mathbf{h}_{\mathrm{init}} - \mathbf{h} \right\| =
    \mathcal{O}\!\left( 
        \sqrt{\frac{BM}{n}}
        \sqrt{\frac{1}{\SNR} + \frac{\| \mathbf{h} \|^2 }{M}}
    \right),
\end{equation}
which shows that the initialisation $\mathbf{h}_{\mathrm{init}}$ becomes more informative as the ratio $BM/n$ decreases.

We note that our setup would be better suited for the scaling regime where the ratio $BM/n$ does not vanish as $n\to\infty$. In that case, the initial channel estimation error $\left\| \mathbf{h}_\text{init} - \mathbf{h} \right\|$ is of the order $\mathcal{O}(1)$ as $n\to\infty$ and $\left\| \mathbf{h}_{\mathrm{t}} - \mathbf{h} \right\|$ decreases with each iteration $t$. In fact, from Remark~\ref{remark_atbt} in the analysis section we get that for $n\to\infty$
\[ 
\left\| \mathbf{h}_{\mathrm{t+1}} - \mathbf{h} \right\|^2 \rightarrow (1-1/\alpha_{t+1})^2\Vert \hv \Vert^2.
\]

\subsection{Improved Initialization Methods}
The initial channel estimate \eqref{eq:h_init} can be further improved. Given that $\Rm:=\Am^\herm \Ym$ contains the desired signal $\xv\hv^\herm$ in Gaussian noise, we can apply a row-wise denoising function to $\Rm$ to get a preliminary estimate of $\xv$. We choose
\begin{equation}
    \hat{\xv}_\ell = \text{softmax}\left[\left(\frac{P}{\tau^2(P + \tau^2)}\|\rv_{l,i}\|^2_2\right)_{i=1,...,B}\right]
\end{equation}
where $\rv_{\ell,i}$ denote the rows of respective section $\Rm_\ell$. The softmax function was defined in \eqref{eq:softmax}, and $\tau^2 = \|\Ym\|_F^2/(nM)$ is the effective noise variance. Then an estimate of $\xv\hv^\herm$ is obtained as
\begin{equation}
    f(\Rm_\ell,\tau^2) = \text{diag}(\hat{\xv}_\ell)\frac{nP}{nP + \tau^2}\Rm_\ell
\end{equation}
This choice is motivated by the fact that $f(\Rm_\ell,\tau^2)$ is the 
posterior mean estimate of an auxiliary signal $\tilde{\Xm}_\ell = \text{diag}(\xv_\ell)\tilde{\Hm}$ in a Gaussian channel $\Rm_\ell = \tilde{\Xm}_\ell + \tau \Wm$ where the matrix $\Hm$ has i.i.d. Gaussian entries. Then the entries $\hat{x}_{\ell,i}$ coincide with the section-wise posterior activity probabilities $\hat{x}_{\ell,i} = P(x_{\ell,i} > 0|\Rm_\ell)$.

This estimator treats the channel vector in each row of $\Rm$ as an independent Gaussian vector, ignoring the fact that they are all identical. In fact, it is the same estimator used to estimate activity probabilities in the MMV-AMP algorithm for a multiple access channel without CSI; see \cite{Liu18, Fen22c}.

Subsequently, an estimate of $\hv$ is obtained by summing $f(\Rm_\ell,\tau)$ over all rows and averaging the results over all sections $l$:
\begin{equation}
    \hv_\text{MMV} = \frac{1}{L}\sum_{\ell=1}^L  \sum_{i=1}^B f(\Rm_\ell,\tau)_i
    \label{eq:h_init_mmv}
\end{equation} 

Another popular choice for finding an estimate of the channel is through a singular value decomposition (SVD), and works as follows: We perform an (economy) SVD of $\Ym$ and set $\hv_\text{SVD} = s_1\vv_1/\sqrt{nP}$ where $s_1$ is the largest singular value and $\vv_1$ is the corresponding right singular vector of $\Ym$. This procedure can recover $\hv$ only up to an overall phase factor $\exp(j\phi)$. 

We find that the estimate obtained through SVD is generally more precise if the missing phase factor can be recovered. This can be done using another estimate $\widehat\hv_2$, which can be either the averaging estimate \eqref{eq:h_init} or the MMV-AMP estimate \eqref{eq:h_init_mmv}.
The missing phase factor can then be found as follows:
\begin{equation}
    \hat{\phi} = \widehat\hv_2^\herm\hv_\text{SVD}
    \label{eq:phi}
\end{equation}
such that the channel estimate is
\begin{equation}
    \hv_\text{init}:= \hat{\phi}\hv_\text{SVD}.
\end{equation}

\section{ Analysis}\label{sec:analysis}

\subsection{Analysis of AMP}
Let $\alpha_0 =\hv^\herm {\hv}_\text{init}/\|{\hv}_{{ \text{init}}}\|^2$ and
$\tau_0^2 = |\alpha_0|^2 P + \sigma^2/{\|{\hv}_{{\text{init}}}\|^2}$.
Then, the SE is given by the following recursive sequence for 
$t\geq 0$:
\begin{subequations}  \label{eq:se}
\begin{align}
\alpha_{t+1} &= \frac{\EE[\|\eta(\Rm,\tau_t)\|^2]}{\EE[\Xm^\top\eta(\Rm,\tau_t)]}\label{at}\\
b_{t+1}^2 &= \frac{\EE[\|\eta(\Rm,\tau_t)\|^2]}{(\sigma^2 + \EE[\|\eta(\Rm,\tau_t)\|^2])^2}\label{bt}\\
\tau_{t+1}^2 &= \frac{\sigma^2\alpha_{t+1}^2}{\|\hv\|^2} +\delta{{\rm mse}}_t(\alpha_t)\label{eq:se_tau}
\end{align}   
\end{subequations}
where, for short, we introduce 
\begin{align}
{\text{mse}}_t(\alpha)&:=   {\frac{1}{N} \EE \left[\|\alpha\Xm - \eta(\Rm,\tau_{t})\|^2\right]}\;.
\label{eq:mse_hat_t}
\end{align}
The expectations in \eqref{eq:se} are taken over $(\Xm,\Rm)$ with $\Xm\sim p_\Xm$ (prior) and $p_{\Rm|\Xm} \sim \mathcal{CN}(\alpha_t\Xm,\tau_t^2\Id_N)$.

Given the SE sequence \eqref{eq:se}, the MSEs of $\xv_{t+1}$ and $\hv_{t+1}$ in Alg. \eqref{eq:amp} are given by 
\begin{align}
\frac{\|\xv-\xv_{t+1}\|^2}{N} &\simeq {\text{mse}}_t(1)\;\label{eq:mse_se}\\
\frac{\|\hv- \hv_{t+1}\|^2}{M} &\simeq \left(\frac{1}{\alpha_{t+1}}-1\right)^2\frac{\|\hv\|^2}{M} + \sigma^2 b_{t+1}^2\;.\label{eq:cmse_se}
\end{align}
Here, and throughout the sequel, we use $\simeq$ as a shorthand notation for convergence in probability. That is, for sequences (in $n$ or $N$) of random vectors $\av,\widehat \av\in \CC^{d}$ we write 
\begin{equation}
 \widehat \av\simeq \av \quad\text{if}\quad \frac{1}{\sqrt{d}}\Vert \widehat \av- \av \Vert\overset{\mathbb P}{\rightarrow} 0 \label{simeq}
\end{equation}
as $n,N\to \infty$ while the aspect ratio $\delta=N/n$ is fixed. $d \in \mathbb{N}_{\geq 0}$ can be arbitrary; we will mainly use $d=1$, $d=n$, or $d=N$).
In Appendix~\ref{sec:simeq}, we collect the key properties of this notation that are used in this paper.

The asymptotic characterisation of the MSE in \eqref{eq:mse_se} and \eqref{eq:cmse_se} follows from Theorem~\ref{Th1} which we state and rigorously prove in Section~\ref{sec:proof}. Our analysis holds for any
Lipschitz continuous $\eta$.
In particular, we do not require that $\eta$ is separable
\footnote{A function $\eta: \CC^N \to \CC^N, \eta(\rv) = [\eta_1(\rv),...,\eta_N(\rv)]^\top$ is called separable if $\eta_i(\rv) = \eta_i(r_i)$.}. We only require that it satisfies Assumption~\ref{as13} in Section~\ref{proof_decop}. Furthermore, Theorem~\ref{Th1} is stated for Gaussian i.i.d. codebooks $\Am$. An extension to the partially orthogonal matrix ensemble \eqref{eq:codebook} is possible but technically involved. We defer it to future work. As discussed later in the comparison of AMP and OAMP, when $\delta$ is fairly large (in this paper we consider $\delta\sim 50$) there is almost no difference between the Gaussian i.i.d. ensemble and \eqref{eq:codebook}. 

A particular difficulty in the analysis of this class of algorithms is that the channel estimates $\hv_t$ may depend on the noise. See, e.g., \cite{Mon21}. However, in the considered scaling regime, that is, fixed $M$, $n,N \to \infty$, all dependencies vanish. In fact, our analysis will show that 
\begin{equation}
    \hv_t \simeq \frac{1}{\alpha_t}\hv + b_t\wv 
    \label{eq:h_hat_gaussian}
\end{equation}
where $\wv$ is independent $\mathcal{CN}(\zerov,\sigma^2\Id_M)$. 

Note that the SE depends on the channel realization through $\Vert \hv\Vert^2$ and $\alpha_0$. Nonetheless, it serves as a tool to quickly simulate the performance of a large number of channel samples. 

\begin{remark}[Estimation of $\alpha_{t+1}$ and $b_{t+1}$] \label{remark_atbt}
To get an algorithmic estimate of $b_t$ during the AMP iterations, we can consider the substitution  $\mathbb{E}\big[\|\eta(\Rm,\tau_t)\|^2\big] \ \longrightarrow \ \|\xv_{t+1}\|^2$ 
which leads to the recursion
\begin{equation}
    b^2_{t+1} = \frac{\|\xv_{t+1}\|^2}{\big(\sigma^2 + \|\xv_{t+1}\|^2\big)^2}.
\end{equation}    
From \eqref{eq:h_hat_gaussian} we have that $\hv_{t} \simeq \hv/\alpha_{t}$ and obtain the simple recursion for $\alpha_t$:
\begin{equation}
    \alpha_{t+1} 
    = \alpha_t \frac{\|\hv_t\|^2}{\hv_{t+1}^\herm \hv_t}, \label{recat}
\end{equation}
which requires only the knowledge of the initial value \(\alpha_0\).
\end{remark}
In principle, the procedure described in Remark \ref{remark_atbt} allows for online estimation of $\alpha_t$ if an accurate estimate of the initial mismatch parameter $\alpha_0$ is available. We do not pursue this direction in this paper, as the proposed algorithms already achieve near-perfect channel estimation.

\subsection{Anaylsis of OAMP}
For the OAMP variation, we have the SE recursion for $t\geq 0$ as

\begin{subequations}\label{eq:oamp_se}
\begin{align}
\alpha_{t+1} &= \frac{\EE[\|{\tilde\eta}(\Rm,\tau_t)\|^2]}{\EE[\Xm^\top{\tilde\eta}(\Rm,\tau_t)]}\\
b_{t+1}^2 &= \frac{\EE[\|\tilde\eta(\Rm,\tau_t)\|^2]}{(\sigma^2 + \EE[\|\tilde\eta(\Rm,\tau_t)\|^2])^2}\\
q_{t+1} &= \EE[\langle\eta'(\Rm,\tau_t)\rangle]\\ 
    \tau_{t+1}^2 &= \frac{\sigma^2\vert \alpha_{t+1}\vert^2}{\|\hv\|^2} + \frac{\delta - 1}{(1-q_{t+1})^2}(\text{mse}_t(\alpha_t) - q_{t+1}^2\tau_t^2)
\end{align}
\end{subequations}
where we introduce the modified denoiser as 
\begin{equation}
\tilde\eta(\Rm,\tau_t)=\frac{\eta(\Rm,\tau_t)-q_{t+1}\Rm}{1-q_{t+1}}\;.
\end{equation}
and expectations are taken over $(\Xm,\Rm)$ with $\Xm\sim p_\Xm$ (prior) and $p_{\Rm|\Xm} \sim \mathcal{CN}(\alpha_t\Xm,\tau_t^2\Id_N)$. 
A formal proof would follow arguments similar to our proof of Theorem~\ref{Th1}, replacing the use of \cite[Theorem~1]{Berthier20} with similar results of \cite{Cad24,ccakmak2025multi}. To avoid repetitive arguments, we omit a formal proof and only give a heuristic derivation of \eqref{eq:oamp_se} in Appendix~\ref{sec:oamp}

\subsection{The SISO component}\label{sec:siso}
\begin{figure}
     \centering
    \subfloat[]{
         \centering
         \includegraphics[width=0.45\textwidth]{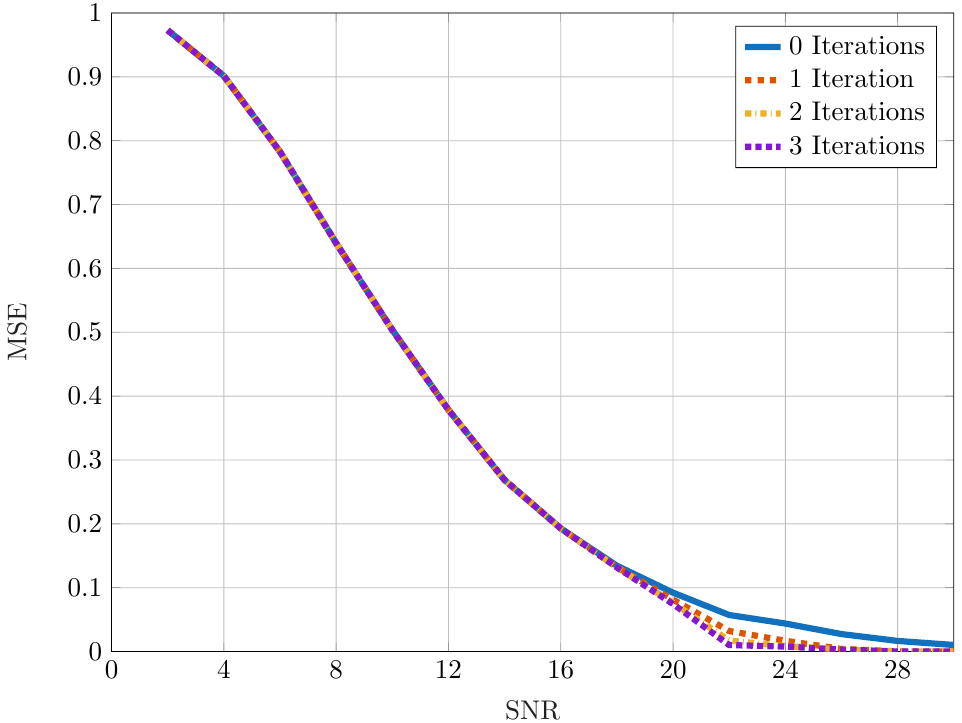}
    }
     \hfill
    \subfloat[Enlarged]{
         \centering
         \includegraphics[width=0.45\textwidth]{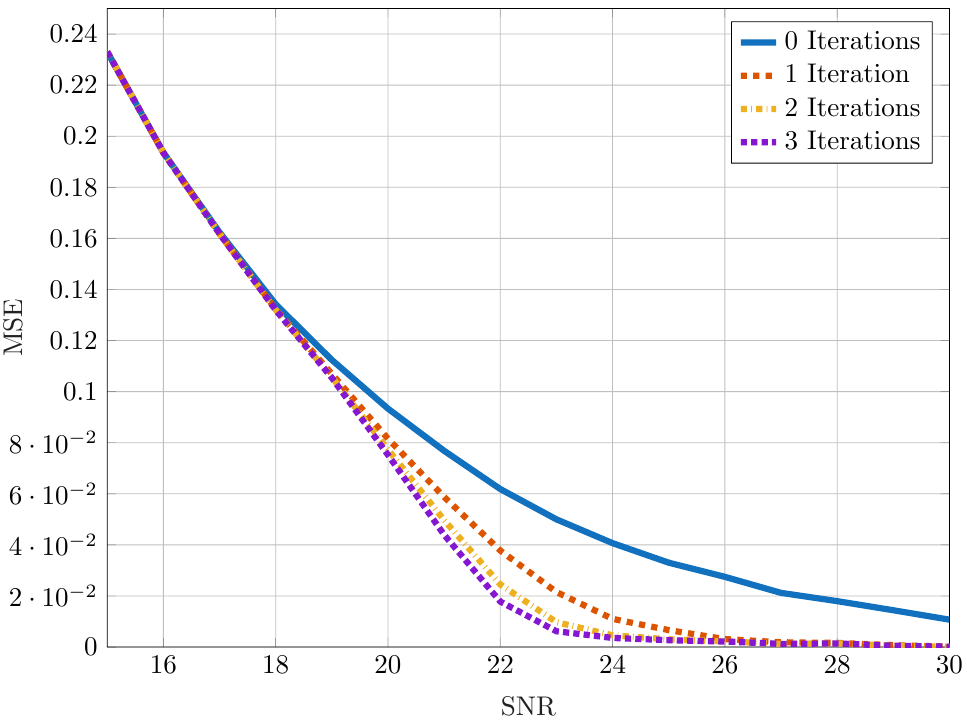}
    }
        \caption{MSE transfer chart of combined demodulation and SISO decoding with a variable number of SISO iterations. }
        \label{fig:mse_transfer}
\end{figure}

To incorporate the outer code into the SE we compute the MSE function of the code empirically. That is, for a sequence of values of $\tau^2$ and $\alpha$ we pre-compute the MSE $\EE[\|\xv - \eta(\rv,\tau)\|]$ of the SISO decoder in an AWGN channel of the form 
\begin{equation}
    \rv = \alpha\xv + \tau\wv  
\end{equation}
for a fixed number of iterations by averaging the MSE over enough samples from the outer codebook. This is in principle possible, but computationally demanding because the parameter space is two dimensional. \footnote{An approximate framework to track the MSE of a SISO decoder of the non-binary LDPC code was given in \cite{Ebe25}. However, as mentioned in \cite{Ebe25} the framework is not always accurate. So for the purpose of predicting the BLER of the SR-LDPC code we resort to empirical pre-computation of the MSE.}
To simplify the evaluation, we first consider the case of known channel vector $
\hat{\hv} = \hv$, in which case $\alpha_t = 1$. We visualise the MSE transfer chart of combined demodulation and SISO decoding in Fig.~\ref{fig:mse_transfer}. It shows the MSE of demodulation and SISO decoding in a channel with $\alpha = 1$ and $\SNR = nLP/\tau^2$. We observe that SISO decoding starts giving an advantage over zero SISO iterations only at high values of the effective SNR. The outer code effectively allows to clean up remaining section errors, once the estimate is good enough. Furthermore, we see that running the SISO decoder for more than one iteration gives only marginal gains. For these reasons, we can minimise decoding complexity by running the SISO decoder for only one iteration at each application of the denoising function, and only when $\tau_t^2$ falls below a certain threshold. The only exception we make is in the last iteration: If the early stoppage criterion described in Section~\ref{sec:decoding} is not met at the last iteration, the SISO decoder is run one last time for $T_\text{BP}$ iterations before creating the final output.

Next, we observe empirically that, in all cases where the decoder is able to reach low residual noise levels, the channel is estimated almost perfectly within a few iterations, that is, $
\alpha_t \to 1$. This can be checked numerically using the SE equations. An important consequence of this is that we can predict the performance of algorithm \eqref{eq:amp} by analysing the case of known $\hv$, i.e., by setting $\alpha_t =1$ in the SE equations. This reduces the complexity of evaluating the SE equations. So, we can pre-compute the SISO MSE curves as a function of only one parameter.  

An additional problem in the given setting is the estimation of the BLER from the effective residual noise variance $\tau_T^2$ at the final iteration of AMP. Given that the vector $\xv$ is chosen from an outer codebook a section error does not necessarily result in a block error. We can pre-compute the BLER for a given $\tau_T^2$ empirically for an AWGN channel. This mapping from $\tau_T$ to BLER can then be combined with SE to predict the overall BLER.

\section{Numerical Results}
\label{sec:results}
For numerical simulations, we use the same outer non-binary LDPC code as in \cite{Ebe25} with $L=766$ sections of size $B = 2^8$ and $n=3675$ complex channel uses.  Of the $L$ sections, $L_\text{inf} = 736$ are information sections, giving an overall coding rate of $R = \log_2 B L_\text{inf}/n \approx 1.6$. We set the maximum number of AMP iterations to $T = 20$. If no valid codeword is found after $T$ iterations, the AMP output is BP decoded one last time for $T_\text{BP} = 10$ iterations before producing the final output. The entries of the fading coefficients and the noise are i.i.d. $\mathcal{CN}(0,1)$.

\subsection{Comparison of Initialization Methods}
In Table~\ref{tab:init_compare} we compare the channel initialisation methods described in Section \ref{sec:init} in terms of channel estimation MSE (CMSE) before and after running AMP/OAMP, as well as the resulting BLER. The values of AMP and OAMP in Table~\ref{tab:init_compare} coincide within the given decimal precision, so we do not distinguish them. For comparison, we also added random initialisation, where $\hv_\text{init}\sim \mathcal{CN}(0,\Id_M)$ and plain SVD, where the missing phase factor, mentioned in Sec.~\ref{sec:init} is not fixed. ``MMV'' denotes estimator \eqref{eq:h_init_mmv}, ``Avg'' denotes estimator \eqref{eq:h_init}, and ``SVD + Avg'', ``SVD + MMV'' denote the SVD estimate where the phase factor has been fixed through \eqref{eq:phi} by one of the other two estimators respectively.

We fix $E_b/N_0 = 1$ dB and set $M=4$. We can observe that SVD + Avg generally gives the best initial channel MSE, but, compared to plain averaging, does not improve the BLER or the final CMSE after the last iteration of AMP decoding. The averaging initialisation \eqref{eq:h_init} is the simplest method that results in optimal BLER. For this reason, we use it exclusively in the remainder of the numerical results section.
\begin{table}
    \centering
    \begin{tabular}{c|ccc}
Initialisation Method & Initial CMSE & Final CMSE & BLER \\
\hline
\hline
Random      & 1.99 & 0.09 & 0.07 \\
SVD         & 2.12 & 0.12 & 0.10 \\
MMV         & 0.19 & 0.00 & 0.05 \\
Avg         & 0.07 & 0.00 & 0.05 \\
SVD + Avg   & 0.03 & 0.00 & 0.05 \\
SVD + MMV   & 0.01 & 0.00 & 0.05 \\
\end{tabular}
    
    \caption{Comparison of Initialization Methods. $E_b/N_0 = 1$ dB, $M = 4$}
    \label{tab:init_compare}
\end{table}

\subsection{State Evolution}
The SE is generally a function of the channel $\hv$. Nonetheless, we can use it to estimate operational quantities like the average BLER by computing the expected value $\EE_\hv[\text{BLER}(\tau_T^2(\hv))]$, where $\tau_T^2(\hv)$ is the residual noise variance at the final iteration $T$ if the channel realisation is $\hv$. Therefore, in the remainder of this subsection, we consider $\hv$ to be fixed but arbitrary. Specifically, for the figures in this subsection, we fix an arbitrary $\hv$ with squared norm $0.9M$ and compute empirical MSE curves by averaging over the distribution of noise and data bits. We choose $M=4$ and $E_b/N_0 = -1$ dB. The decoder is run without the outer BP decoder, since we are interested in the dependence of the decoder on the channel estimate and the behaviour of the BP decoder does not depend on the channel estimate.

\begin{figure}
    \centering
    \subfloat[$\Delta = 1.1\exp(i\pi/16)$]{
         \centering
         \includegraphics[width=0.45\textwidth]{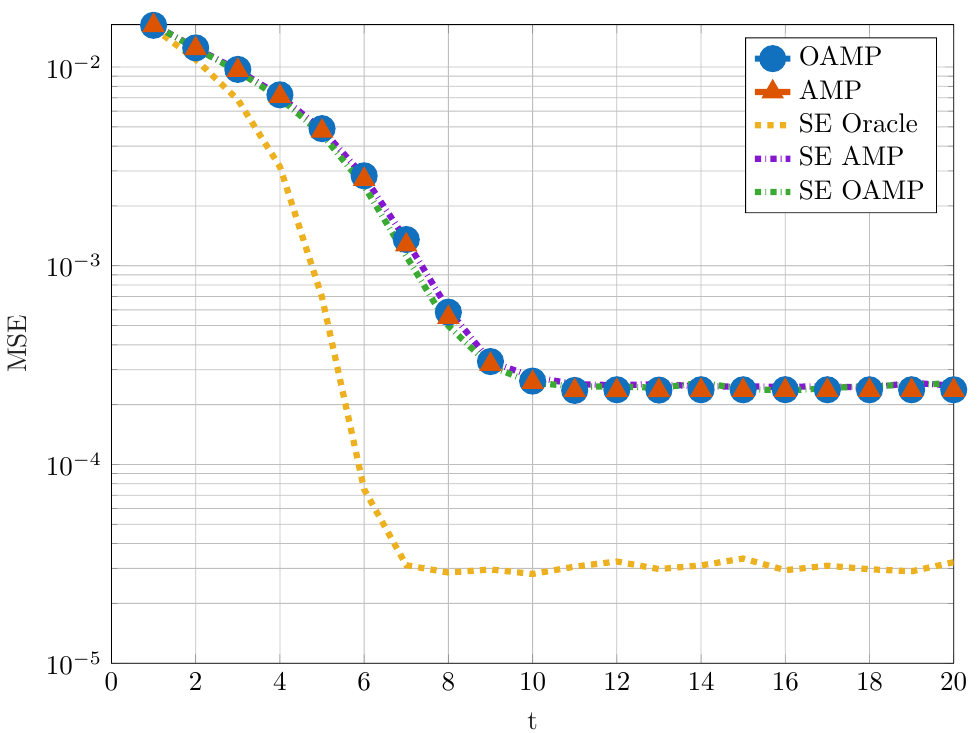}
     }
    \hfill
    \subfloat[$\Delta = 2.1\exp(3.5i\pi/16)$]{
         \centering
         \includegraphics[width=0.45\textwidth]{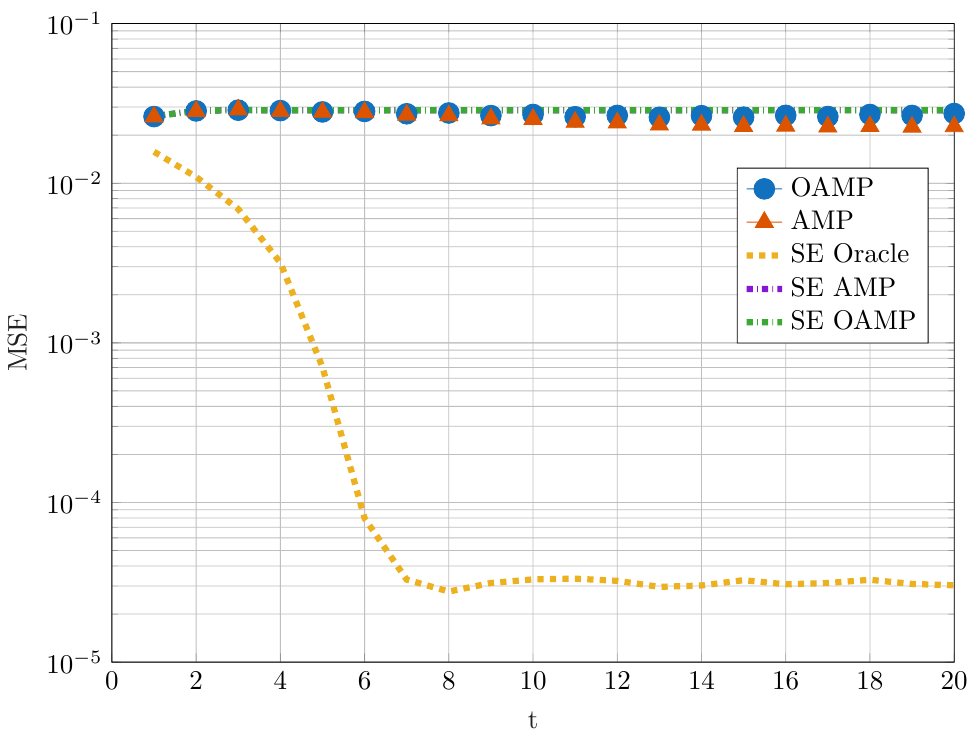}
    }
     \caption{State evolution for a fixed mismatched channel estimate $\hat{\hv} = \Delta\hv$.}
        \label{fig:MSE_h_delta}
\end{figure}

\begin{figure}
     \centering
     \subfloat[MSE of $\xv_t$]{
         \centering
         \includegraphics[width=0.45\textwidth]{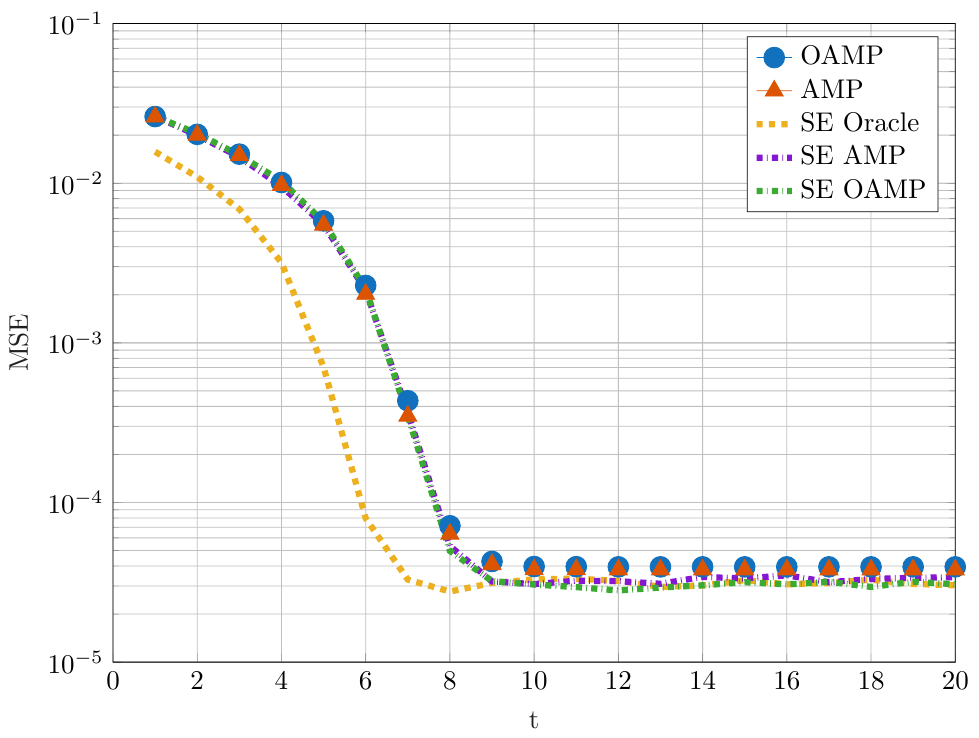}
     }
     \hfill
     \subfloat[Channel estimation MSE]{
         \centering
         \includegraphics[width=0.45\textwidth]{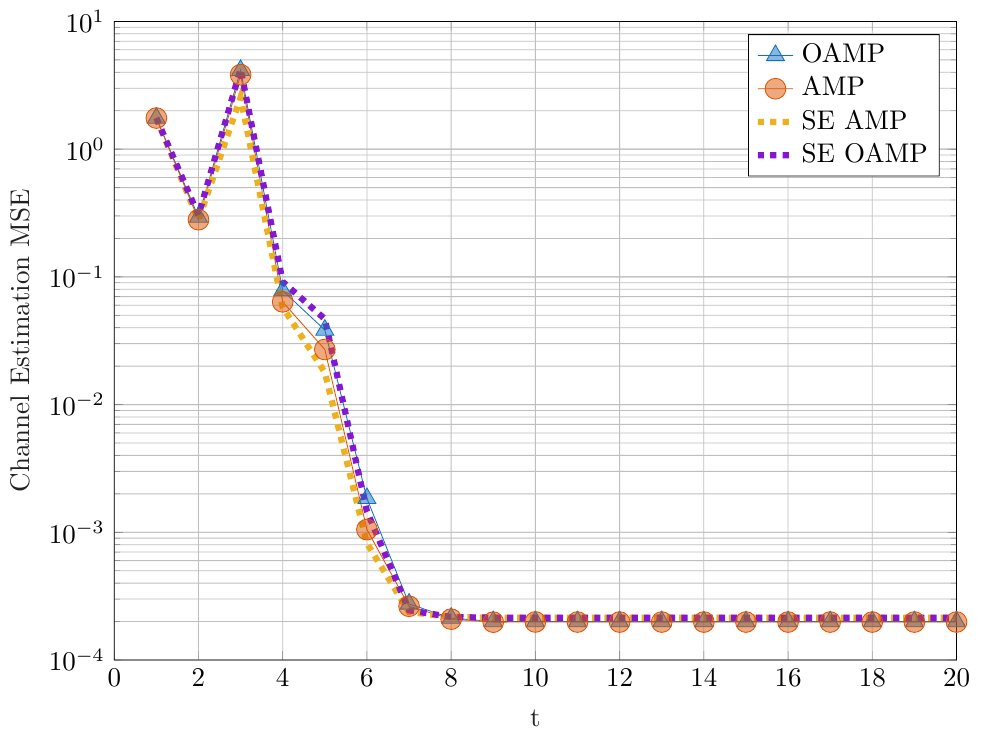}
     }
    \caption{State evolution for the initialisation $\hv_0 = 2.1\exp(3.5i\pi/16)\hv$ with updated channel estimates.}
        \label{fig:MSE_h_delta_est}
\end{figure}

To evaluate the precision of the given analysis for mismatched channel estimates, we first consider the case where $\hat\hv := \hv_t = \text{const}.$ stays fixed during iterations and is set to $\hat{\hv} = \Delta\cdot\hv$ for some fixed $\Delta\in \CC$. We compare the performance and SE prediction for the AMP and OAMP based algorithms \eqref{eq:amp} and \eqref{eq:oamp} respectively and refer to them as AMP and OAMP. 
In the left of Fig.~\ref{fig:MSE_h_delta} we choose $\Delta = 1.1\exp(i\frac{\pi}{16})$ and in the right $\Delta = 2.1\exp(3.5i\frac{\pi}{16})$. 
However, in Fig.~\ref{fig:MSE_h_delta_est}, we keep the same initial value of $\hat{\hv}$ as in Fig.~\ref{fig:MSE_h_delta} (right), but the channel estimates are updated according to \eqref{eq:h_hat}. The corresponding channel estimation MSE is shown on the right of Fig.~\ref{fig:MSE_h_delta_est}. The value of $\Delta$ here leads to an artificially low $|\alpha_0|$ which is never observed with the initialisation \eqref{eq:h_init}. For such a high offset we can observe that the oracle MSE is not reached any more and the CMSE exhibits a non-monotonic behaviour. However, our SE analysis can accurately predict both effects. 

In general, the initial channel estimate $\hv_\text{init}$ is a function of the noise realisation $\Wm$. Hence, the SE prediction also becomes a function of $\Wm$ through $\alpha_0(\hv,\hv_\text{init})$. Nonetheless, we observe a good concentration of the average MSE around the SE prediction. For the SE curves, the exact same channel initialisations were used as for the empirical MSE.

\begin{figure}
     \centering
     \subfloat[MSE of $\hat{\sv}$]{
         \centering
         \includegraphics[width=0.45\textwidth]{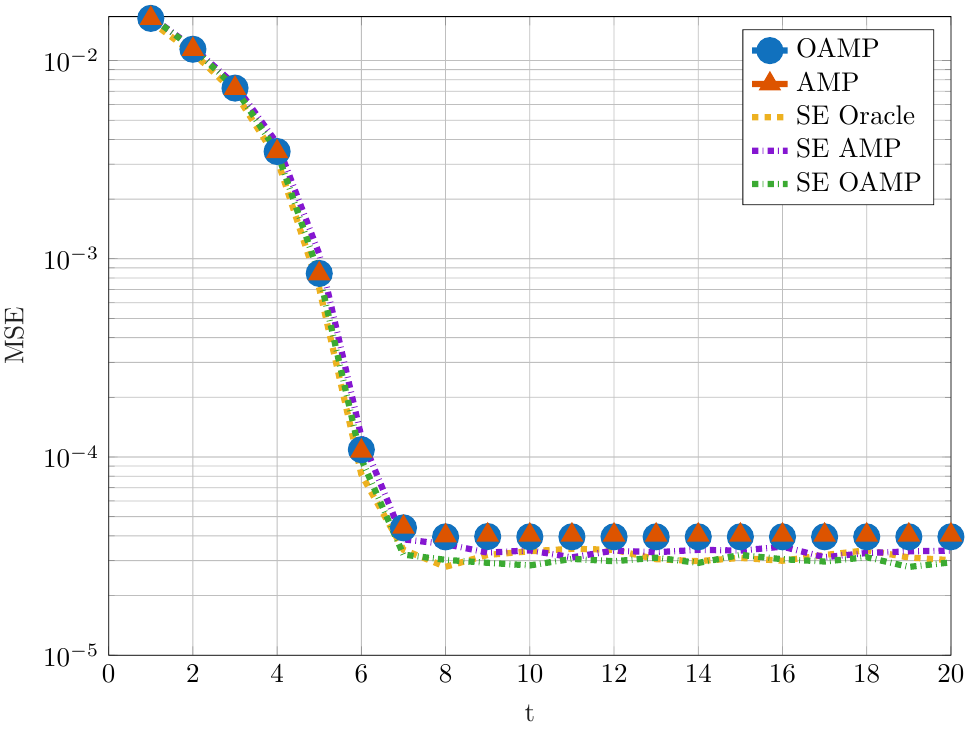}
    }
     \hfill
     \subfloat[Channel estimation MSE]{
         \centering
         \includegraphics[width=0.45\textwidth]{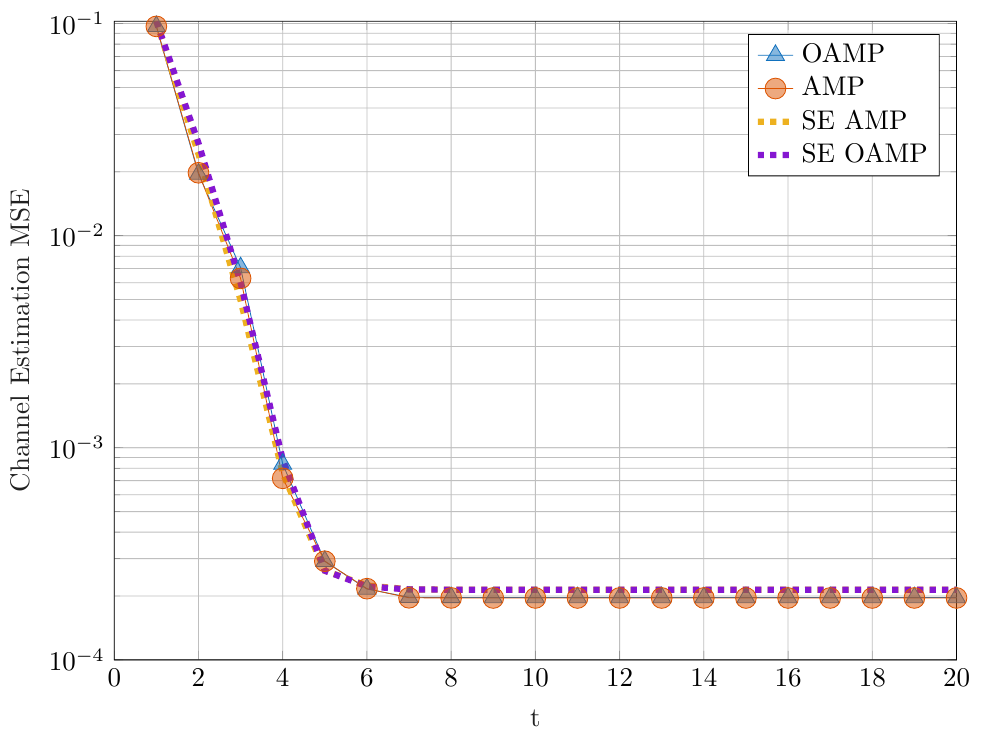}
    }
        \caption{State evolution for initialisation via \eqref{eq:h_init} with updated channel estimates.}
        \label{fig:se_with_init}
\end{figure}

With re-introduced channel estimation, both algorithms are well described by the SE given in \eqref{eq:se} and coincide with the SE with full CSI. This is visualised in Fig.~\ref{fig:se_with_init}.

In most figures in this section, we observe very little difference in performance between AMP and OAMP. This is expected since
the SE analysis shows that for the case of full CSI at the receiver the effective noise variance should differ between AMP and OAMP by a factor $\frac{\delta - 1}{\delta}$. For the considered code $\delta \approx 50$, and that factor becomes very small. Only when close to a phase transition we expect the difference to become noticeable.

\begin{figure}[ht]
    \centering
    \includegraphics[width=0.45\textwidth]{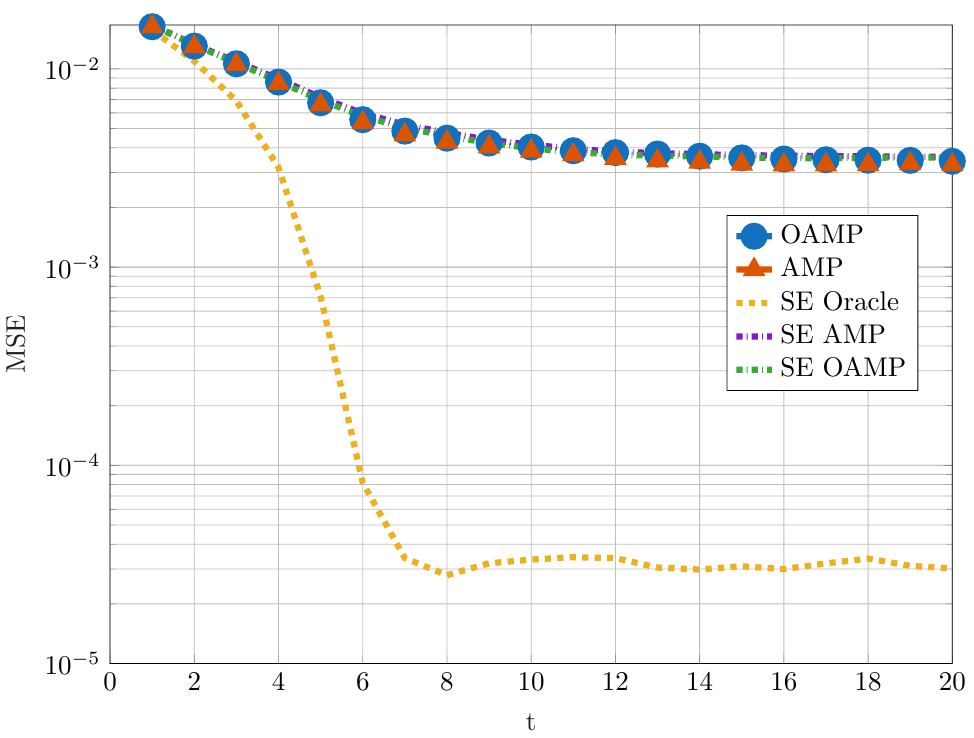}
    \caption{State evolution for a mismatched channel estimate chosen as $\hat{\hv} = \hv_\text{init}$ from \eqref{eq:h_init}.}
    \label{fig:MSE_h_avg_fixed}
\end{figure}

\subsection{Block Error Rate}
For the simulations in this section, AMP and OAMP perform identically, so we do not distinguish them. For the simulations in this section, we averaged the BLER over $
hv$ by Monte-Carlo simulations.

\begin{figure}[ht]
    \centering
    \includegraphics[width=0.6\textwidth]{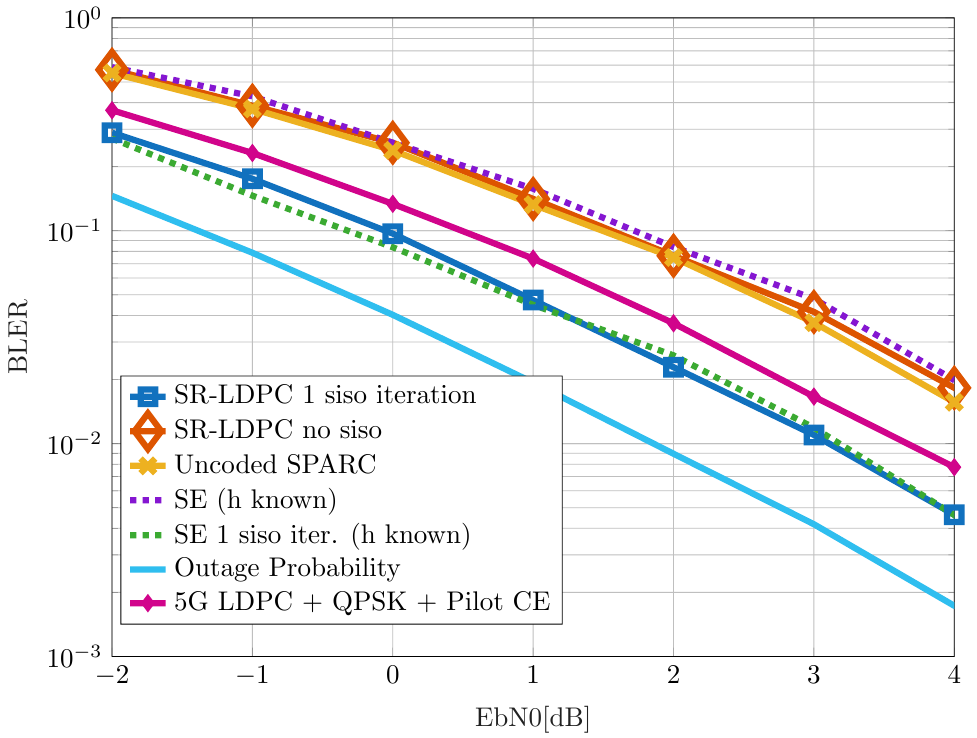}
    \caption{Comparison of the proposed coding scheme with and without outer SISO decoding, uncoded SR codes, and the 5G LDPC code with QPSK modulation and pilot based channel estimates. $M=4$}
    \label{fig:5G}
\end{figure}

In Fig.~\ref{fig:5G}, we compare the BLER of the AMP/OAMP decoder without SISO BP decoding with the AMP decoder with one BP iteration. 
The decoder without BP decoding emulates the strategy in \cite{Gre18} where outer decoding is only performed after the final AMP iteration. Both decoders employ the initialisation strategy described in Section \ref{sec:init}. The results show that adding a single SISO BP iteration leads to a gain of about 2 dB. Furthermore, we compare the results to the oracle SE prediction, with full CSI at the receiver, created as described in Section \ref{sec:analysis}. We observe a full overlap between SE and empirical results, confirming that in this setting our decoder can reach the same performance as a decoder with full CSI. We omitted empirical results for the decoder with full CSI because they match the curves without CSI perfectly.
We also plot the BLER of uncoded SPARCs with the same rate, where we set $L = 736$ and no outer code is used. We can see that if no SISO iterations are used, there is no coding gain over the uncoded version, while the use of the outer code improves performance by almost 2dB.

In Fig.~\ref{fig:5G} we also compare the BLER of the proposed coding scheme with the 5G LDPC code (MATLAB implementation from the 5G toolbox, base graph 2 and CRC attachment according to the standard) with QPSK modulation and pilot-based channel estimation. The total blocklength is set to $n=3750$ complex symbols, of which $n_p$ are used to transmit pilot symbols. The number $n_p$ was empirically optimised and set to $n_p = 50$. We choose $M=4$ receive antennas. The number of transmitted bits is set to $k=5864$ to match the rate of the SR-LDPC code. 

In addition, we compare the BLER of the coding schemes to the outage probability $p_\text{out} = \mathbb{P}(R < \log(1 + \|\hv\|^2\text{SNR})) = \mathbb{P}(\|\hv\|^2 < \frac{2^R - 1}{\text{SNR}})$ which serves as achievability bound. The rate $R$ is set to the coding rate of the SR-LDPC code. Technically, the bound is asymptotic and assumes full CSI. A finite-length bound for the case without CSI is available in \cite{Yan14}. However, since the dispersion of the quasi-static Rayleigh fading channel is zero \cite{Yan14} the convergence to the asymptotic limit is very fast and we find that the finite-length bound from \cite{Yan14} gives no noticeable change to the outage probability for the considered block length of $n= 3675$.

\begin{figure}[ht]
    \centering
    \includegraphics[width=0.6\textwidth]{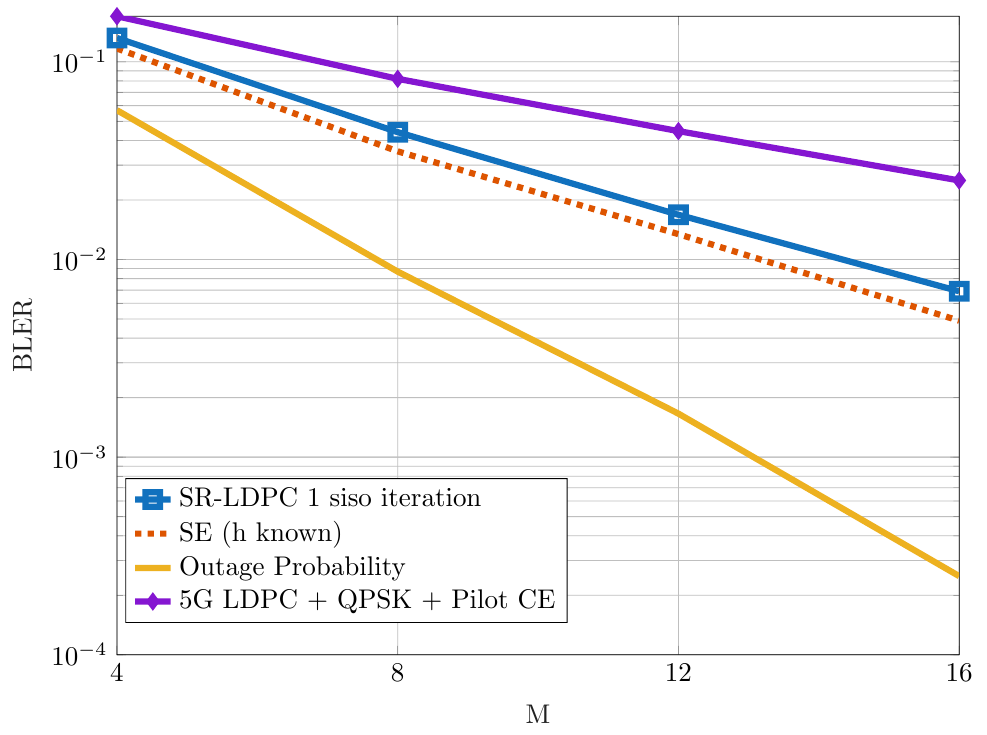}
    \caption{Comparison of the proposed coding scheme with the 5G LDPC code, QPSK modulation and pilot based channel estimates. $E_b/N_0 = 5.5$dB, $\hv$ normalised by $M$.}
    \label{fig:5G_M}
\end{figure}

In Fig.~\ref{fig:5G_M} we plot the BLER as a function of the number of receive antennas $M$. To make the BLER comparable for different $M$, for this figure only, we normalise the channel norm by $M$. In this way, we observe only the pure diversity gain as a function of $M$. For the 5G LDPC coding scheme, the number of pilot symbols is emprically optimised and set to $n_p = 75$ for $M=4$ and $n_p = 100$ for $M>=8$. We again observe a gain of the SR-LDPC over the 5G LDPC code, which increases with $M$. We can also see that the gap to the outage probability of both schemes increases with $M$, showing potential for improvements. As $E_b/N_0$ increases further, our experiments (not shown here) confirm that the slopes of the outage probability of the different schemes approach the same value, as expected from an analysis of the diversity-multiplexing trade-off in the high-SNR regime \cite{Tse04}. 
 

\section{The Proof of the SE Analysis of AMP}
\label{sec:proof}
In this section, we provide a rigorous high-dimensional analysis of the AMP algorithm \eqref{eq:amp}. To streamline the proof, we replace the algorithmic updates of $\tau_t$ and $q_{t+1}$ with their theoretical counterparts from the SE \eqref{eq:se}. That is, we set $q_{t+1} =\EE[\langle\eta'(\Rm,\tau_t)\rangle]$ and $\tau_t$ as in \eqref{eq:se_tau}. Such a substitution is common in the analysis of AMP algorithms and can be justified inductively across iterations (see also \cite[Corollary~2]{Berthier20}).

Also, for readability, we defer certain technical assumptions (specifically Assumption~\ref{as13}) to Appendix~\ref{proof_decop}.     
\begin{theorem}[Decoupling principle]\label{Th1}
Let $A_{ij}\sim_{\text{i.i.d.}}\mathcal {CN}(0,1/n)$ and let  Assumption~\ref{as13} hold. Then, for any deterministic sequences  $T_N:(\CC^{N})^2\to \CC$ and $T_n:(\CC^{n})^{2}\to \CC$ of uniformly pseudo-Lipschitz test functions of order $k$ (see \cite[Section 3.1]{Berthier20}) we have for any $0\leq t\leq T$
\begin{subequations}
\begin{align}
T_N(\xv,\rv_t)      &\simeq T_N(\xv,\alpha_t\xv+\tau_{t}\zv_N)\label{rv1_th1}\\
T_n(\zv,\pv_{t+1})  &\simeq T_n(\zv,m_{t+1}\zv+\tau_{\phi,t+1}\zv_n)\label{pv1_th1}\\
\hv_{t+1}           &\simeq \frac{1}{\alpha_{t+1}}\hv+b_{t+1}\wv
\end{align}    
\label{decoup_th1}
\end{subequations}
where  $\zv_N\sim \mathcal{CN}(\zerov, \Id_N)$, $\zv_n\sim \mathcal{CN}(\zerov, \Id_n)$, and $\wv\sim\mathcal{CN}(\zerov, \sigma^2\Id_M)$ is an independent projection of $\Wm$ into an $M$-dim subspace (specifically, $\wv=\Wm^\herm\vv$, where $\vv\in\CC^L$ is independent of $\Wm$ and $\norm{\vv}=1$) and $\zv:= \Am\xv$.
The dynamical order parameters $b_{t+1}$, $\alpha_t$ and $\tau_t$ 
for $t\geq 0$ are all defined as in SE \eqref{eq:se}. Moreover, we construct for $t\geq 0$
\begin{subequations}
 \begin{align}
m_{t+1}&=\frac{\mathbb E[\Xm^\top\eta(\alpha_{t+1}\Xm+\tau_{t}\Zm,\tau_{t})]}{\sqrt{N}\norm{\xv}}\\
\tau_{\phi,t+1}^2&= \frac{\delta}{N}\mathbb E[\norm{\eta(\alpha_{t}\Xm+\tau_{t}\Zm,\tau_t)]}^2]-\delta m_{t+1}^2\;.
\end{align}   
\end{subequations}
\end{theorem}
We recall the notion of concentration in terms of $\simeq$ notation introduced in \eqref{simeq}. We also refer to Appendix~\ref{sec:simeq} where we present its elementary properties, which are frequently used in the following proof.

We emphasise that Theorem~\ref{Th1} is a general result that holds for any sequence of Lipschitz denoisers $\eta$ and any distribution of $\xv$ that satisfies
Assumption~\ref{as13}. In particular, Assumption~\ref{as13} is fulfilled for the case where $\xv$ is a codeword from an outer NB-LDPC code and $\eta$, as described in Section~\ref{sec:denoising}, contains a SISO BP decoder that is run for fewer iterations than the girth of the NB-LDPC code. In fact, we show that a stronger statement holds: The SE analysis is independent of the particular transmitted codeword. Thus, for the analysis, we can assume without loss of generality that the all-zero codeword was transmitted. Although this is expected from the geometric uniformity property shown in \cite{Ebe25}, we explicitly verify it in Appendix~\ref{sec:allzero}.
\begin{remark}
    We note that, in principle, it is possible to strengthen the notion of convergence in Theorem~\ref{Th1} from convergence in probability to almost sure convergence by invoking recent results from \cite[Theorem~1]{reeves2025dimension}. However, bypassing some of the assumptions, like positive-definiteness of the $T\times T$ joint state-evolution matrix (i.e., $\Sigma$ in \cite[Theorem~1]{reeves2025dimension}) requires (relatively) lengthy arguments (see perturbation approach in \cite[Section~5.4]{Berthier20}), which are outside the scope of this paper.  
\end{remark}
\subsection*{The proof of Theorem~\ref{Th1}}
For convenience we substitute $\sigma^2\to \delta\sigma^2/(LP)$ so that 
\begin{equation}
\frac{\norm{\xv}}{\sqrt N}=1. 
\end{equation}
The key element in our argument is that, without loss of generality, we can assume that
\begin{equation}
\Am = \sqrt{\delta}\frac{\uv\xv^\top}{N} +\widetilde \Am \Pm_{\xv}^\perp \label{H0}
\end{equation}
where we define
\begin{equation}
    \Pm_{\xv}^\perp = \Id- \frac{\xv\xv^\top}{N}\; ,
\end{equation}
$\uv\sim\mathcal{CN}(\mathbf 0,\Id_n)$, and $\widetilde\Am \in \CC^{n\times N}$ with $\widetilde A_{ij}\sim_{\text{i.i.d.}}\mathcal{CN}(0,1/n)$ independent of $\uv$. Conditioned on $\xv$, we have $A_{ij}\sim_{\text{i.i.d.}}\mathcal{CN}(0,1/n)$, and thereby $\Am$ and $\xv$ are independent. Note from \eqref{H0} that 
\begin{equation}
\zv=\Am\xv=\sqrt{\delta}\uv \;.\label{Ym}
\end{equation}

At a high level, we employ the representation \eqref{H0} to decompose the AMP dynamics into the random vectors $\uv$ and $\xv$, together with a new AMP dynamics driven by the random matrix $\widetilde{\Am}$. This decomposition allows us to conveniently treat the channel estimation mismatch. Our SE analysis is then verified by inductively invoking previous results established for generic AMP iterations with non-separable denoisers \cite{Berthier20}.

We begin with using \eqref{H0} to represent the original AMP dynamics \eqref{eq:amp} in terms of $\widetilde{\Am}$:
\begin{subequations}
 \begin{align}
\widetilde\xv_t&=\Pm_{\xv}^\perp\xv_t\\
\uv_t&=\frac{\Wm\hv_{t}}{\Vert\hv_t \Vert^2}+\sqrt{\delta}
\left(\frac{\hv^\herm{\hv}_t}{\Vert \hv_t\Vert^2}
-\langle \xv,\xv_t\rangle\right)\uv-\phiv_t\\
\psiv_t&= \widetilde\Am^\herm\uv_t+ \widetilde\xv_t\\
\phiv_{t+1}&=\widetilde\Am \widetilde\xv_{t+1}-\delta q_{t+1}\uv_t\;
\end{align}   
\end{subequations}
where we set $\phiv_0=\zerov$ and, also, for convenience, we express the dynamics separately as
\begin{align}
\rv_t&= \left(\frac{\langle \uv,\uv_t\rangle}{\sqrt{\delta}}+\langle\xv,\xv_t\rangle\right)\xv + \Pm_{\xv}^\perp \psiv_t\\
\pv_{t+1}&=\sqrt{\delta}\langle\xv,\xv_{t+1}\rangle\uv+\phiv_{t+1}\;.\label{pvt}
\end{align}
Here, and throughout the sequel, we denote for $\av, \bv \in \mathbb{C}^d$
\begin{equation}
\langle \av, \bv \rangle:= \frac{1}{d} \av^\herm \bv\;.
\label{eq:inner_prod}
\end{equation}

Now, let $\mathcal H_{t'}$ stand for the hypothesis that for all $0\leq t\leq t'$ we have 
\begin{subequations}
\label{Hp}
\begin{align}
\langle \xv,\xv_t\rangle&\simeq m_{t} \label{m0}\\
\hv_t&\simeq \frac{\hv}{\alpha_t}+b_{t}\wv\label{alpha0}\\
\langle \uv,\uv_t\rangle&\simeq\sqrt{\delta}(\alpha_t-m_t)\label{uz0}\;
\end{align}  
\end{subequations}
where we set $m_0=0$ and $b_{0}=0$. Note that the initial hypothesis $\mathcal H_0$ is evident.

Following the notation of \cite{Berthier20}, we introduce the functions for all $t\geq 0$  
\begin{align}
g_{t}(\phiv)&:=  \alpha_t\frac{\Wm\hv}{\norm{\hv}^2}+ \sqrt{\delta}(\alpha_t - m_t)\uv-\phiv \\
e_t(\psiv)&:= \eta(\alpha_t\xv+\Pm_\xv^\perp\psiv,\tau_t)-m_{t+1}\xv\;.
\end{align}
Note that both functions $g_{t}$ and $e_{t}$ are Lipschitz, because $\eta(\cdot, \tau_t)$ is Lipschitz. Hence, this implies that 
$e_t(\xv + {\epsilonv}) \simeq e_t(\xv)$ for 
${\epsilonv} \simeq \zerov$. Then, hypothesis $\mathcal{H}_{t'}$ implies 
that for any $0 \leq t \leq t'$ 
\begin{subequations}
\label{Hpc}
\begin{align}
\uv_{t}&\simeq g_{t}(\phiv_t)\label{uvt}\\
\widetilde\xv_{t+1}&\simeq e_t(\psiv_t)\label{rv0}\;.
\end{align}    
\end{subequations}
Here, we note that the dynamical order parameter $b_t$ scales as $1/\sqrt{n}$, i.e., $b_t \wv \simeq \zerov$.  Indeed, it will become clear later that $b_t$ is a finite-sample correction parameter. %

Next, we verify from \eqref{Hpc} that for $0\leq t\leq t'$
\begin{subequations}
\label{effecfields}
 \begin{align}
\psiv_t&\simeq \widetilde \psiv_t\\
\phiv_{t+1}&\simeq \widetilde \phiv_{t+1}
\end{align}  
\end{subequations}
where $\widetilde \psiv_t$ and $\widetilde \phiv_{t+1}$ are constructed from the AMP dynamics for $0\leq t\leq t'$ according to 
\begin{subequations}
\label{eq:amp_berth}
 \begin{align}
\widetilde\psiv_t&=\widetilde\Am^\herm g_{t}(\widetilde\phiv_t)+ e_{t-1}(\widetilde\psiv_{t-1})\\
\widetilde\phiv_{t+1}&=\widetilde\Am\ev_t(\widetilde\psiv_t)-\delta q_{t+1}g_{t}(\widetilde\phiv_t)\;
\end{align} 
\end{subequations}with $\widetilde\phiv_0=\zerov$ and $e_{-1}(\cdot)=\mathbf 0$. We verify the asymptotic equivalence \eqref{effecfields} recursively: Let $\psiv_{t-1}\simeq \widetilde \psiv_{t-1}$ and $\phiv_{t}\simeq \widetilde \phiv_{t}$. Then, from the Lipschitz property of $g_t$ together with \eqref{uvt} we get
\begin{align}
\uv_{t}&\simeq g_{t}(\phiv_{t})\simeq g_{t}(\widetilde\phiv_{t})\;.
\end{align}  
Furthermore, we have 
\begin{align}
\psiv_t&\simeq \widetilde\Am^\herm g_{t}(\widetilde\phiv_{t})+ e_{t-1}(\widetilde\psiv_{t-1})\;.
\end{align}
Here, we use of the Lipschitz property of $e_{t-1}$ and $g_t$ and together with the property (i) in Appendix~\ref{sec:simeq}. Similarly, from the Lipschitz property of $e_t$ together with \eqref{rv0} we have 
\begin{align}
\tilde\xv_{t+1}&\simeq e_{t}(\widetilde\phiv_{t})\;.
\end{align} 
Then, by the Lipschitz property of $g_{t}$ and $e_t$ and the property (i) in Appendix~\ref{sec:simeq}, we get 
\begin{equation}
\phiv_{t+1}\simeq\widetilde\Am\ev_t(\widetilde\psiv_t)-\delta q_{t+1}g_{t}(\widetilde\phiv_t)\;.
\end{equation}

The form of \eqref{eq:amp_berth} coincides with the generic AMP recursion studied in \cite{Berthier20}. In particular, from \cite[Theorem~1]{Berthier20} we verify in Appendix~\ref{proof_decop} the following modified \emph{decoupling principle} for all $0\leq t\leq t'$:  for any deterministic sequences  $T_N:(\CC^{N})^2\to \CC$ and $T_n:(\CC^{n})^{M+2}\to \CC$ of uniformly pseudo-Lipschitz test functions of order $k$ we have
\begin{subequations}
\label{decoup}
\begin{align}
T_N(\xv,\rv_t)&\simeq T_N(\xv,\alpha_t\xv+\tau_{t}\zv_N)\label{rv1}\\
T_n(\Wm,\uv,\phiv_{t+1})&\simeq T_n(\Wm,\uv,\tau_{\phi,t+1}\zv_n)\label{pv1}\;.
\end{align}    
\end{subequations}
Notice that the above decoupling principle is slightly different from that in Theorem~\ref {Th1}. To show the equivalence (w.r.t. the notion $\simeq$), we first note that $\langle\xv,\eta(\rv,\tau)\rangle$ is a pseudo-Lipschitz function of $(\xv,\rv)$ of order $2$. Hence, from \eqref{rv0} and \eqref{rv1} we get 
\begin{align}
\langle\xv,\xv_{t'+1} \rangle\simeq m_{t'+1}\;.\label{mv1}
\end{align}
Furthermore, from \eqref{pvt}  this implies for all $0\leq t\leq t'$ 
\begin{equation}
\pv_{t+1}\simeq\sqrt{\delta}m_{t+1}\uv +\phiv_{t+1}\label{pv2}\;.
\end{equation}
Then, from the property (ii) in the Appendix~\ref{sec:simeq} we get
\begin{align}
T_n(\Wm,\uv,\pv_{t+1})&\simeq T_n(\Wm,\uv,\sqrt{\delta}m_{t+1}\uv +\phiv_{t+1})\\&=:  \widetilde T_{n}(\Wm,\uv,\phiv_{t+1})
\end{align}
where the mapping $\widetilde T_{n}:(\CC^{n})^{M+2}\to \CC$ is uniformly pseudo-Lipschitz of order $k$. 
This establishes the equivalence of \eqref{decoup} with Theorem~\ref{Th1}.

We complete the proof by verifying $\mathcal{H}_{t'+1}$ from \eqref{decoup}, as this will then imply that the decoupling principle \eqref{Hpc} holds for $t = t' + 1$. Note that we have already verified that \eqref{m0} holds for $t=t'+1$ (see \eqref{mv1}). Then, we next verify \eqref{alpha0} and \eqref{uz0} for $t=t'+1$:
From \eqref{Ym} we note that
\begin{equation}
    \hv_{t+1}
    = \frac{ \sqrt{\delta}\langle \uv,\pv_{t+1}\rangle}
    { \tfrac{\sigma^2}{n} + \tfrac{1}{n}\|\pv_{t+1}\|^2 }\hv+\frac{ 
    \tfrac{1}{n}\Wm^\herm \pv_{t+1} }{ \tfrac{\sigma^2}{n} + \tfrac{1}{n}\|\pv_{t+1}\|^2 }.
\end{equation}
Then, using \eqref{pv2} together with \eqref{pv1} and property (iii) in the Appendix~\ref{sec:simeq} we get 
\begin{align}
\sqrt{\delta}\langle \uv,{\pv}_{t'+1}\rangle&\simeq\delta m_{t'+1}\\
\frac{1}{n}\|\pv_{t'+1}\|^2&\simeq \delta m_{t'+1}^2+\tau_{\phi, t'+1}^2\;.
\end{align}
Although from \eqref{pv1} it follows that  $\tfrac{1}{n}\Wm^\herm \pv_{t+1}\simeq \zerov$, we can refine this and represent it as Gaussian noise with variance scaling as $1/n$.  Specifically,  we note that $f_n(\xv)=\frac{1}{\sqrt n
}\Wm^\herm\xv$ is uniformly Lipschitz as $\norm{\frac{1}{\sqrt n}\Wm}_2$ a.s. bounded as $n\to\infty$. We then get from \eqref{pv1} 
\begin{align}
    \frac{1}{\sqrt n}\Wm^\herm\pv_{t'+1}
    &\;\simeq\;\frac{1}{\sqrt n}\Wm^\herm\left(\sqrt{\delta}\,m_{t'+1}\uv+\tau_{\phi,t'+1}\zv_n\right) \\
    &\;=\;\sqrt{\delta m_{t'+1}^2+\tau_{\phi,t'+1}^2}\,\frac{\|\tilde\zv\|}{\sqrt n}\,
        \frac{\Wm^\herm\tilde\zv}{\|\tilde\zv\|} \\
    &\;\simeq\;\sqrt{\delta m_{t'+1}^2+\tau_{\phi,t'+1}^2}\,\wv
\end{align}
where $\tilde\zv\sim\mathcal{CN}(\zerov,\Id_n)$ is independent of $\Wm$, and $\wv:= \frac{\Wm^\herm\tilde\zv}{\|\tilde\zv\|}\sim \mathcal{CN}(\zerov,\sigma^2\Id_M)$ with nothing that $\wv$ is independent of $\tilde \zv$. The latter step follows from the fact that $\frac{\|\tilde\zv\|}{\sqrt n}\simeq 1$. Hence, we have 
\begin{equation}
    \hv_{t'+1}\;\simeq\; \frac{1}{\alpha_{t'+1}}\hv \;+\; b_{t'+1}\wv \;.\label{new}
\end{equation}
We also get from this result that
\begin{equation}
 \uv_{t'+1}\simeq \alpha_t\frac{\Wm\hv}{\norm{\hv}^2} +\sqrt{\delta}(\alpha_{t'+1}-m_{t'+1})\uv+\phiv_{t+1}\;.\label{uvt2}
\end{equation}
Then, using \eqref{uvt2} together with \eqref{pv1} yields that \eqref{uz0} holds for $t = t'+ 1$. 

\section{Summary}
\label{sec:summary}
In this paper, we present an AMP decoding algorithm for a concatenated coding scheme consisting of SPARCs and an outer NB-LDPC code for a slow fading Rayleigh channel without CSI at the receiver. We develop an analytical framework to predict the BLER of the coding scheme together with the channel estimation error. The framework is asymptotically rigorous. Numerical results confirm the accuracy of our analytic framework for finite blocklengths and show that the coding scheme can achieve the same performance as if full CSI were available, with only minimal overhead in computational complexity. The results show that SPARCs can achieve competitive performance on fading channels, compared to standardised coded modulation solutions with pilot-based channel estimation. Our framework can serve as a tool for further optimisation and the extension to more complex channel models, like the multi-user setting.  

\newpage
\appendices
\section{Preliminaries}
\label{sec:simeq}
Here, we present several elementary properties of the notion of concentration in terms of the notation $\simeq$ as defined in \eqref{simeq}. For convenience, we set $d\in\{n,N\}$ and $\Bm\in\{\frac{1}{\sqrt n }\Wm, \Am,\Am^\herm\}$ and a vector $\epsilonv\simeq \zerov$ of appropriate dimension. We have the following properties: 
\begin{itemize}
    \item [(i)] We have $\Bm\epsilonv\simeq \zerov$.
    \item [(ii)] Consider a random vector $\av\in \CC^{d}$ such that $\frac{\norm{\av}}{\sqrt d}\simeq p $ for a constant $p$. Then, for any deterministic sequences  $T_d:\CC^{d}\to \CC$ of uniformly pseudo-Lipschitz test functions of order $k$, we have  $T_d(\av+\epsilonv)\simeq T_d(\av)$. 
    \item [(iii)] Consider random vectors $\av,\bv\in \CC^{d}$ where  $ \av\sim_\text{i.i.d.} {A}$ and  $\bv\sim_\text{i.i.d.} B$  with $\mathbb E[\vert A\vert^2]<\infty$ and $\mathbb E[\vert B\vert^2]<\infty$. Then, for any $\widehat{\av}\simeq\av$ and ${\widehat \bv}\simeq \bv$ we have $\frac{1}{d}\widehat{\av}^\herm \widehat{\bv}\simeq \mathbb E[\bar AB]$.
\end{itemize}
The proof of (i) follows from the inequality  
$ \norm{\Bm \epsilonv} \leq \norm{\Bm}_2 \, \norm{\epsilonv}$  
where $\norm{\Bm}_2$ denotes the largest singular value of $\Bm$, which is a.s. bounded as $d \to \infty$. The proof of (ii) follows directly from the definition of pseudo-Lipschitz mappings (see \cite[Eq.~(20)]{Berthier20}). As to the proof of (iii), let  
$\epsilonv_a := \widehat{\av} - \av$ and $\epsilonv_b := \widehat{\bv} - \bv$ and define  
\begin{align}
\Delta := \widehat{\av}^\herm \widehat{\bv} - \av^\herm \bv 
= \av^\herm \epsilonv_b + \bv^\herm \epsilonv_a + \epsilonv_a^\herm \epsilonv_b. \label{decom2}
\end{align}  
By the law of large numbers,  
$
\frac{\norm{\av}}{\sqrt{d}} \simeq \sqrt{\mathbb{E}[|A|^2]}$ and
$\frac{\norm{\bv}}{\sqrt{d}} \simeq \sqrt{\mathbb{E}[|B|^2]}  
$. Hence, $\Delta/d \simeq 0$ which implies 
\begin{equation}
\frac{1}{d}\widehat{\av}^\herm \widehat{\bv} 
\simeq \frac{1}{d}\av^\herm \bv 
\simeq \mathbb{E}[\bar{A}B]
\end{equation}  
where the last step again follows from the law of large numbers.

\section{Proof of Remark~1}\label{prem1}
In expressing $\Ym^\herm\Am\mathds{1}/\Vert\xv \Vert_{1}$, we can rescale $\xv$ and $\Wm$ such that $\Wm\to\Wm/\sqrt{nP}$ and $\xv\in\{0,1\}^{N}$.  Then $\Vert\xv \Vert_{1}=\Vert\xv \Vert^2=L$. To simplify the dependence between $\Ym$ and $\Am\mathds{1}$, we introduce the orthonormal vectors
\begin{subequations}
 \begin{align}
\vv_{0}&=\frac{\xv}{\sqrt{L}}\label{v0}\\
\vv_{1}&=\frac{\mathds{1}-\xv}{\sqrt{(B-1)L}}\;.\label{v1}
\end{align}   
\end{subequations}
Without loss of generality, we can assume that $\Am$ is constructed as 
\begin{align}
\Am\equiv\gv_{0}\vv_{0}^\top+\gv_{1}\vv_{1}^\top
+\widetilde \Am\left[\Id_N -\sum_{i=0,1}\vv_{i}\vv_i^\top\right]\;,\label{rep}
\end{align}
where $\gv_{0},\gv_{1}\sim\mathcal {CN}(\mathbf 0,\Id_{n}/n)$ and $\widetilde \Am\sim \Am$ is a matrix from the Gaussian ensemble. Indeed, it is easy to verify that conditioned on $\vv_{0}$ and $\vv_{1}$, we have $A_{ij}\sim_{\text{i.i.d}}\mathcal{CN}(0,1/n)$; thus $\Am$ is statistically independent of $\vv_{0}$ and $\vv_{1}$. 

We note from \eqref{v0} and \eqref{v1} that $\mathds{1}= \sqrt{L}\vv_0+\sqrt{(B-1)L}\vv_1$. Then, we get from \eqref{rep} that
\begin{subequations}
 \begin{align}
\frac{\Ym}{\sqrt{\Vert\xv \Vert_1}}&=\gv_0\hv^\herm+ \sqrt{\frac{1}{nPL}}\Wm\\
\frac{\Am\mathds{1}}{\sqrt{\Vert\xv \Vert_1}}&=\gv_0+ \sqrt{B-1}\gv_1\;.
\end{align}   
\end{subequations}
This completes the proof.
\section{The Proof of the modified decoupling principle}\label{proof_decop}
Here we prove \eqref{decoup}.
We recall that we assume $A_{ij}\sim_{\text{i.i.d.}}\mathcal {CN}(0,1/n)$. Then, for our system model, all the premises of \cite[Theorem~1]{Berthier20} are fulfilled under the following assumption. 
\begin{assumption}\label{as13} 
Let $\Xm_N$ (the index $N$ is dropped for readability) be a sequence of random variables taking values in some set $\mathcal{X}_N \subset \CC^N$. For each $t,s\geq 0$ and for any two jointly Gaussian random vectors $\Psim_1,\Psim_2\in \CC^{N}$ with $[\Psim_1^\top, \Psim_2^\top]^\top \sim \mathcal{CN}(\zerov,\mathcal{C}\otimes \Id_N)$ and some arbitrary covariance matrix $\mathcal{C}\in \CC^{2 \times 2}$, there exist deterministic constants $m,c\in\RR$ such that for each $\xv \in \mathcal{X}_N$ as $N\to\infty$
\begin{subequations}
 \begin{align}
 \mathbb E[\langle\xv,\eta(\alpha_t\xv+\Psim_1,\tau_t)\rangle\mid\xv] &\overset{\PP}{\rightarrow} m\label{eq:as1_condition1}\\
\mathbb E[\langle\eta(\alpha_t\xv+\Psim_1,\tau_t),\eta(\alpha_s\xv+\Psim_2,\tau_s) \rangle\mid \xv] &\overset{\PP}{\rightarrow} c\;.\label{eq:as1_condition2}
\end{align}  
\label{eq:as1_conditions}
\end{subequations}

\end{assumption}
We recall from \eqref{eq:inner_prod} that $\langle 
\cdot,\cdot\rangle$ denotes the normalised inner product.
Note also that the entries of both the random vector $\Xm$ and the random vector $\eta(\alpha_t\Xm + \Psim_1, \tau_t)$ are bounded by $1$. Therefore, by the uniform integrability, the above convergences in probability imply convergences in expectation, i.e,  
\begin{subequations}
\label{sep}
\begin{align}
\mathbb E[\langle\Xm,\eta(\alpha_t\Xm+\Psim_1,\tau_t)\rangle]&= m +o(1)\\
\mathbb E[\langle\eta(\alpha_t\Xm+\Psim_1,\tau_t),\eta(\alpha_s\Xm+\Psim_2,\tau_s) \rangle] &= c+o(1)\;.
\end{align}
\end{subequations}
where $o(1)$ denotes a deterministic sequence (of $n$ or $N$) such that $o(1)\to 0$ as $n,N\to \infty$. 

From \cite[Theorem~1]{Berthier20} we have for all $0\leq t\leq t'$:
\begin{subequations}
\label{bertier20}
\begin{align}
T_N(\xv,\rv_t)&\simeq \mathbb E[T_N(\xv,\alpha_t\xv+\tilde\tau_{t}\Pm_{\xv}^\perp\Zm_N)\mid \xv]\\
T_n(\Wm,\uv,\phiv_{t+1})&\simeq \mathbb E[T_n(\Wm,\uv,\tilde\tau_{\phi,t+1}\Zm_n)\mid \Wm,\uv]
\end{align}    
\end{subequations}
where the variances are constructed as  
\begin{align}
\tilde\tau_{t}^2&=\lim_{n\to \infty}\frac{1}{n}\mathbb E\left[\norm{g_t(\tilde\tau_{\phi,t}\Zm_n)}^2\mid \Wm,\uv\right]\label{exist1} \\
&\overset{\rm a.s}{=}\lim_{n\to \infty}\frac{1}{n}\mathbb E[\norm{g_t(\tilde\tau_{\phi,t}\Zm_n)}^2]\label{exist11}
\\
\tilde\tau_{\phi,t+1}^2&=\lim_{N\to \infty}\frac{\delta}{N}\mathbb E\left[\norm{e_t(\tilde\tau_{t}\Zm_N)}^2\mid \xv\right]\label{exist2}\\
&\overset{\PP}{=}\lim_{N\to \infty}\frac{\delta}{N}\mathbb E[\norm{e_t(\tilde\tau_{t}\Zm_N)}^2]\;.\label{exist22}
\end{align}
The existence of the limit \eqref{exist1} and the self-averaging property (holding a.s.) \eqref{exist11} follow from the fact that $g_t(\cdot)$ is a linear mapping and $W_{ij} \sim_{\text{i.i.d.}} \mathcal{CN}(0,\sigma^2)$ and $\uv \sim \mathcal{CN}(\zerov,\Id_n)$. Furthermore, the existence of the limit in \eqref{exist2} follows from Assumption~\ref{as13}, while the self-averaging property (holding in probability) \eqref{exist22} follows from \eqref{sep}. Note that the results \eqref{bertier20} are valid for any fixed $\Wm,\uv,\xv$ (under Assumption~\ref{as13}).

From the Gaussian concentration of pseudo-Lipschitz mappings (see, \cite[Lemma~C.8]{Berthier20}) we note e.g. 
\begin{equation}    
T_N(\xv,\alpha_t\xv+\tilde\tau_{t}\Pm_{\xv}^\perp\zv_n)\simeq \mathbb E_{\Zm_N}[T_N(\xv,\alpha_t\xv+\tilde\tau_{t}\Pm_{\xv}^\perp\Zm_N)\mid \xv]\;.
\end{equation}
In other words, we have 
\begin{subequations}
\label{ePm}
\begin{align}
T_N(\xv,\rv_t)&\simeq T_N(\xv,\alpha_t\xv+\tilde\tau_{t}\Pm_{\xv}^\perp\zv_N)\\
T_n(\Wm,\uv,\phiv_{t+1})&\simeq T_n(\Wm,\uv,\tilde\tau_{\phi,t+1}\zv_n)\;.
\end{align}    
\end{subequations}
Next, we eliminate $\Pm_\xv^\perp$ in \eqref{ePm}. Specifically, since $\Pm_\xv^\perp \zv_N \simeq \zv_N$, it follows from the property (ii) in the Appendix~\ref{sec:simeq} that
\begin{equation}
T_N(\xv,\alpha_t\xv+\tilde\tau_{t}\Pm_{\xv}^\perp\zv_N)\simeq T_N(\xv,\alpha_t\xv+\tilde\tau_{t}\zv_n)\;.
\end{equation}
Thus, we are left with verifying that for all $0\leq t\leq t'$
\begin{subequations}
\begin{align}
T_N(\xv,\alpha_t\xv+\tilde\tau_{t}\zv_N)&\simeq T_N(\xv,\alpha_t\xv+\tau_{t}\zv_N)\\
T_n(\Wm,\uv,\tilde\tau_{\phi,t+1}\zv_n)&\simeq T_n(\Wm,\uv,\tau_{\phi,t+1}\zv_n)\;.
\end{align}    
\end{subequations}
From the property (ii) in the Appendix~\ref{sec:simeq}, we solely need to verify that for all $0\leq t\leq t'$
\begin{align}
\tilde\tau_{t}&\simeq\tau_{t}\\
\tilde\tau_{\phi,t+1}&\simeq\tau_{\phi,t+1}\;.
\end{align}
To this end, we use the Lipschitz property of $e_t$ and $g_t$ together with invoking the reverse triangle inequality
and get the bounds for a constant $C_t$
\begin{align}
\left|\frac{1}{n}\Vert g_t(\tilde a\uv) \Vert^2-\frac{1}{n}\Vert g_t(a\uv) \Vert^2\right|&\leq  C_t \frac{\norm{\uv}^2}{n} 
(\vert a \vert+\vert\tilde a\vert)\vert \tilde a-a\vert\label{Boundn}\\
\left|\frac{1}{N}\Vert e_t(\tilde a\uv) \Vert^2-\frac{1}{N}\Vert e_t(a\uv) \Vert^2\right|&\leq C_t \frac{\norm{\uv}^2}{N} 
(\vert a \vert+\vert\tilde a\vert)\vert \tilde a-a\vert\label{BoundN}\;.
\end{align}
Thus, if $a\simeq\tilde a$ and $\uv\sim\mathcal {CN}(\zerov,\Id)$, then the bounds converge to zero (in probability) as $n,N\to \infty$. Now suppose $\tilde\tau_{\phi,t}\simeq \tau_{\phi,t}$. Then, from \eqref{BoundN} we get $\tilde\tau_{t}\simeq \tau_{t}$. Then, using this result together with \eqref{BoundN} we get  $\tilde\tau_{\phi,t+1}\simeq \tau_{\phi,t+1}$.  This completes the proof.

\section{Reduction to the all-zero codeword}
\label{sec:allzero}

Here, we show that when $\xv$ is an arbitrary codeword from an NB-LDPC code, and $\eta_t$ is the denoiser as described in Section~\ref{sec:denoising}, the quantities on the left side of \eqref{eq:as1_conditions} in Assumption~\ref{as13} are equal for all codewords $\xv$. Thus, without loss of generality, we can assume in the SE analysis that $\xv$ is the all-zero codeword (that is, each section has a non-zero only in the first position). With slight abuse of notation, we use the natural identification $\FF_B \to \{0,1\}^B, q\mapsto \lambda_q$, to identify a codeword $\xv\in\FF_B^L$ with its sparse representation $\xv\in\{0,1\}^{BL}$.
It should be clear from the context to what representation we refer to.

Let $\cv$ denote a codeword and let $P_\cv$ denote a diagonal block matrix $\text{diag}(\Cm_1,...,\Cm_L)$ where each $\Cm_i$ is a permutation matrix that represents a circular shift by $c_i$. We first show the following invariance property for an arbitrary vector $\rv\in\CC^N$, an arbitrary $\tau > 0$, and a denoiser $\eta(\rv)=\eta(\rv,\tau)$ as described in Section~\ref{sec:denoising}.
\begin{equation}
    \Pm_\cv \eta(\rv) = \eta(\Pm_\cv \rv) 
    \label{eq:eta_permutation}
\end{equation}
Since $\eta$ is a concatenation of multiple functions, we show the property for each factor. The property holds for $\eta_1$ in \eqref{eq:eta1} since $\eta_1$ is invariant under any section-wise permutation, in particular under circular shifts. In addition, the variable-to-check messages \eqref{eq:bp_vntocn}, as well as the final output \eqref{eq:bp_out}, are invariant under arbitrary section-wise permutations. It remains to show the invariance property for the check-to-variable messages \eqref{eq:bp_cntovn}.    
Recall that $\mu_{a\to i}(q)$, as well as $\mu_{i\to a}(q),\ q\in\FF_B$, represent probability distributions on $\xv_i \in \FF_B$. Therefore, the action of $\Cm_i$ on $\mu_{a \to i}(q)$ is $\tilde\mu_{a \to i}(q) := \Cm_i\mu_{a \to i}(q) = \mu_{a \to i}(q+c_i)$. 
\begin{align}
    &\sum_{\substack{\qv'\in \FF_B^{|\partial a\setminus i|}\\\Hm_{a,\partial a\setminus i}\qv' = - H_{a,i}q}} \prod_{j \in \partial a\setminus i} \tilde\mu_{j\to a}(q'_j) \\
    &= \sum_{\substack{\qv'\in \FF_B^{|\partial a\setminus i|}\\\Hm_{a,\partial a\setminus i}\qv' = - H_{a,i}q}} \prod_{j \in \partial a\setminus i} \mu_{j\to a}(q'_j+c_j)\\
    &= 
    \sum_{\substack{\qv'\in \FF_B^{|\partial a\setminus i|}\\\Hm_{a,\partial a\setminus i}\qv' = - H_{a,i}(q+c_i)}} \prod_{j \in \partial a\setminus i} \mu_{j\to a}(q'_j)\\
    &\sim \tilde\mu_{a \to i}(q)
\end{align}
where in the second to last line we used that $\Hm\cv = \zerov$, and therefore $\Hm_{a,\partial a\setminus i}\cv_{\partial a\setminus i} = - H_{a,i}c_i$. This completes the proof of \eqref{eq:eta_permutation}.

With \eqref{eq:eta_permutation}, it is straightforward to prove that the conditions in Assumption~\ref{as13} are fulfilled. Let $\xv$ be an arbitrary codeword, let $\xv_0$ denote the representation of the all-zero codeword, and choose $\cv = -\xv$. Then 
\begin{align}
    \EE_{\Psim_1}\langle\xv,\eta(\alpha_t\xv + \Psim_1,\tau_t)\rangle 
    &= \EE_{\Psim_1}\langle\Pm_{-\xv}\xv,\Pm_{-\xv}\eta(\alpha_t\xv + \Psim_1,\tau_t)\rangle\\
    &= \EE_{\Psim_1}\langle\xv_0,\eta(\alpha_t\xv_0 + \Pm_{-\xv}\Psim_1,\tau_t)\rangle\\
    &= \EE_{\Psim_1}\langle\xv_0,\eta(\alpha_t\xv_0 + \Psim_1,\tau_t)\rangle
\end{align}
where in the first line we used that $\Pm_\xv$ is an orthogonal matrix for any $\xv$, in the second line we used \eqref{eq:eta_permutation}, and in the last line we used that the distribution of $\Psi_1$ is invariant under arbitrary section-wise permutations. The same steps can be used to show that the left side of \eqref{eq:as1_condition2} is equal for all $\xv$.

\section{Heuristic Derivation of the OAMP SE}
\label{sec:oamp}
Consider the iterations
\begin{subequations}
\begin{align}
    \rv_t &= \sv_t + \Am^\herm\left(\frac{\Ym \hv_t}{\|\hv_t\|^2} - \Am \sv_t\right)\\
    \sv_t &= \tilde{\eta}_t(\rv_t)
\end{align}
\label{eq:oamp_short}
\end{subequations}
with $\sv_0 = \zerov$, some initial channel estimate $\hv_0$, and $\Ym$ given by \eqref{eq:outer1}. $\tilde{\eta}_t$ is chosen as 
\begin{equation}
    \tilde{\eta}_t(\rv) = \frac{1}{1-q_t}(\eta(\rv,\tau_t) - \rv q_t)
    \label{eq:eta_div_free}
\end{equation}
with
\begin{equation}
    q_t = \langle \eta'(\rv,\tau_t)\rangle
\end{equation}
and $\eta(\rv,\tau_t)$ is the same denoising function \eqref{eq:eta} as in AMP \eqref{eq:amp}.
Then $\tilde{\eta}_t$ is approximately divergence-free, that is, $\EE[\tilde{\eta}_t'(\Rm)] = 0$.
With this choice of $\tilde{\eta}_t$, equations \eqref{eq:oamp_short} become \eqref{eq:oamp}.

To derive the SE, we define the error vectors $\zetav_t := \rv_t - \alpha_t\xv$ and $\fv_t := \sv_t - \alpha_t\xv$, where $\alpha_t = \hv^\herm \hv_t/\|\hv_t\|^2$. We get from \eqref{eq:oamp_short} that
\begin{subequations}
\begin{align}
    \zetav_t &= \Bm \fv_t + \Am^\herm\Wm \frac{\hv_t}{\|\hv_t\|^2}\\
    \fv_{t+1} &= \tilde{\eta}_t(\alpha_t\xv + \zetav_t) - \alpha_t\xv.
\end{align}
\end{subequations}
Here $\Bm = \Id_N - \Am^\herm\Am$.
Following the argumentation in \cite{Ma17}, under the usual assumptions of OAMP the variances $\nu_t^2:= \EE[\|\fv_t\|^2]/N$ and $\tau_t^2:=\EE[\|\zetav_t\|]^2/N$ satisfy the recursion
\begin{subequations}
\begin{align}
    \nu_t^2 &= \frac{1}{N}\EE[\|\tilde{\eta}_t(\alpha_t\xv + \tau_t\Zm) - \alpha_t\xv\|^2]\\
    \tau_{t+1}^2 &= \frac{1}{N}\trace(\Bm\Bm^\herm)\nu_t^2 + \frac{\sigma^2}{\|\hat{\hv}\|^2}.
\end{align}
\end{subequations}
For partial DFT matrices $\trace(\Bm\Bm^\herm)/N = \delta - 1$. Furthermore, expanding the definition of $\tilde{\eta}_t$ in \eqref{eq:eta_div_free} and defining $\text{mse}_t(\alpha)$ as in \eqref{eq:mse_hat_t}, we get the following.
\begin{equation}
    \tau_{t+1}^2 = \frac{\sigma^2}{\|\hv_t\|^2} + \frac{\delta - 1}{(1-q)^2}\left(\text{mse}_t(\alpha_t) - q^2\tau_t^2\right)
    \label{eq:oamp_se1}
\end{equation}
At convergence, i.e., when $q\to 0$, we see that \eqref{eq:oamp_se1} differs from \eqref{eq:se_tau} only in the factor before $\text{mse}_t(\alpha_t)$, which is $(\delta-1)$ for OAMP, coming from the partial orthogonal matrix structure, and $\delta$ for GAMP, which implicitly assumes Gaussian matrices.

The rationale behind this choice of $\tau_t$ in \eqref{eq:oamp_tau} is that $\|\uv_t\|^2/n$ converges to $\sigma^2/\|\hv_t\|^2 + \delta \nu_t^2$ as dimensions grow large. Plugging this into \eqref{eq:oamp_tau} gives that $\|\uv_t\|^2/n$ is asymptotically equal to $\sigma^2/\|\hv_t\|^2 + (\delta - 1)\nu_t^2$ and therefore provides an estimate of $\tau_{t}^2$. 
This results precisely in the joint iterations \eqref{eq:oamp_se}.

\printbibliography
\end{document}